\documentclass[twocolumn]{aastex7}
\usepackage{natbib}

\usepackage{subfig}

\usepackage{amsmath}

\usepackage{listings}

\usepackage{xcolor}

\usepackage[T1]{fontenc}

\definecolor{codegreen}{rgb}{0,0.6,0}
\definecolor{codegray}{rgb}{0.5,0.5,0.5}
\definecolor{codepurple}{rgb}{0.58,0,0.82}
\definecolor{backcolour}{rgb}{0.95,0.95,0.92}

\lstdefinestyle{mystyle}{
    backgroundcolor=\color{white},   
    commentstyle=\color{black},
    keywordstyle=\color{black},
    numberstyle=\tiny\color{black},
    stringstyle=\color{black},
    basicstyle=\ttfamily\footnotesize,
    breakatwhitespace=false,         
    breaklines=true,                 
    captionpos=b,                    
    keepspaces=true,                                   
    numbersep=5pt,                  
    showspaces=false,                
    showstringspaces=false,
    showtabs=false,                  
    tabsize=2,
    basicstyle=\ttfamily,
  upquote=true,
  columns=fullflexible,
  keepspaces=true,
  showstringspaces=false,
  breaklines=true, 
}

\begin{document}

\title{A Probabilistic Framework for Population Studies of the Solar Neighborhood:\\Application to SDSS-V and Gaia}

\author[0000-0003-3410-5794]{Ilija Medan}
\affiliation{Department of Physics and Astronomy,
	Vanderbilt University,
	Nashville, TN 37235, USA}
\affiliation{Canadian Institute for Theoretical Astrophysics, University of Toronto, Toronto, ON M5S-98H, Canada}
\email{ilija.medan@vanderbilt.edu}

\author[0000-0002-3481-9052]{Keivan G. Stassun}
\affiliation{Department of Physics and Astronomy,
	Vanderbilt University,
	Nashville, TN 37235, USA}
\email{keivan.stassun@vanderbilt.edu}

\author[0000-0003-0179-9662]{Zachary Way}
\affiliation{Department of Physics and Astronomy, Georgia State University, Atlanta, GA 30302, USA}
\email{zway1@gsu.edu}

\author[0000-0003-1479-3059]
{Guy S. Stringfellow}
\affiliation{University of Colorado Boulder, Boulder, Colorado 80309, USA}
\email{Guy.Stringfellow@colorado.edu}

\author[0000-0002-1379-4204]{Alexandre Roman-Lopes}
\affiliation{Department of Astronomy, Universidad de La Serena, Av. Raul Bitran 1302, La Serena, Chile}
\email{aroman@userena.cl}

\author[0000-0002-3651-5482]{Jiadong Li}
\affiliation{Max-Planck-Institut für Astronomie, Königstuhl 17, D-69117 Heidelberg, Germany}
\email{jdli@mpia.de}

\author[0000-0001-7297-8508]{Madeline Lucey}
\affiliation{
Department of Physics \& Astronomy, University of Pennsylvania, 209 S 33rd St., Philadelphia, PA 19104, USA }
\email{madlucey@sas.upenn.edu}

\author[0000-0003-0174-0564]{Andrew R. Casey}
\affiliation{School of Physics and Astronomy, Monash University, Clayton VIC 3800, Australia}
\affiliation{Center for Computational Astrophysics, Flatiron Institute, 162 Fifth Avenue, New York, NY 10010, USA}
\email{andrew.casey@monash.edu}

\author[0000-0002-0149-1302]{B\'{a}rbara Rojas-Ayala}
\affiliation{Instituto de Alta Investigaci\'on, Universidad de Tarapac\'a, Casilla 7D, Arica, Chile}
\email{brojasayala@academicos.uta.cl}

\author[0000-0002-7795-0018]{Ricardo L\'opez-Valdivia}
\affil{Universidad Nacional Aut\'onoma de M\'exico, Instituto de Astronom\'ia, AP 106, Ensenada 22800, BC, M\'exico}
\email{rlopezv@astro.unam.mx}

\author[]{Jos\'e G. Fern\'andez-Trincado}
\affil{Centro de investigaci\'on en Astronom\'ia, Facultad de Ingenier\'ia, Ciencia y Tecnolog\'ia, Universidad Bernardo O’Higgins, Av. Viel 1497, Santiago, 8370993, Chile}
\email{jose.fernandez@ucn.cl}

\begin{abstract}
Studies of the Solar Neighborhood require spectroscopic follow-up of stars identified in astrometric surveys to fully characterize their physical properties. The SDSS-V Solar Neighborhood Census (SNC) is a dedicated program to observe stars within 100~pc. However, due to competing observing programs and fiber assignment constraints, the resulting sample carries severe and complex selection effects. A framework is presented for characterizing the selection function of the SDSS-V SNC relative to the Gaia Catalog of Nearby Stars (GCNS), along with a forward modeling method to infer the properties of stellar subpopulations across the  GCNS-defined 100 pc sample. The selection function is based on a method that models the selection probability as a function of sky position, Gaia $G$ magnitude, and $\added{BP-RP}$. The resulting detection probabilities faithfully reproduce the known survey planning logic. This work further introduces the concept of a ``subpopulation probability"---a grid of posterior estimates across the Hertzsprung-Russell (HR) diagram representing the likelihood that a GCNS member belongs to a given SDSS-V defined subpopulation. The framework is validated with a mock dataset and its scientific utility is demonstrated through two applications using data from the Data Release 19: mapping H$\alpha$ emission across the HR diagram and measuring the variation of stellar \added{density} with \added{mass and} metallicity. These results illustrate how statistically robust population studies can be conducted with an incomplete spectroscopic survey when the selection function is well characterized. The code is made publicly available with this work, which will serve as an important tool for future studies.
\end{abstract}


\section{Introduction}\label{sec:intro}

Studies of the Solar Neighborhood ($< 100$ pc) are crucial for understanding of stellar populations across the full stellar mass range. Critically, a census of the Solar Neighborhood allows for the unique observation of the lowest mass M dwarfs, and even brown dwarfs, which are too faint to observe at larger volumes. Observations of low-mass M dwarfs are critical for studies of stellar populations, as they account for three out of every four stars \citep{henry2006, RECONS}. When these low-mass stars can be placed in a three-dimensional volume, it opens the door to understanding them as a population, such that their true density as a function of, e.g.~metallicity, age and Galactic population can be determined.

Over the past few decades, there has been significant progress toward obtaining a complete census of the local population of stars. This has been done primarily through dedicated astrometric surveys, the most complete of which are the 10 pc sample from REsearch Consortium On Nearby Stars \citep[RECONS;][]{RECONS}, the 20 pc sample from \citet{kirkpatrick2024}\added{ and the 25 pc from of the Fifth Catalogue of Nearby Stars \citep[CNS5;][]{CNS5}}. These samples are statistically complete and provide strong constraints on the mass and luminosity function of their constituents. While these nearby-volume samples are complete, they lack the volume to probe rare stellar types (e.g., O/B stars). Gaia now provides the necessary larger volume while maintaining high completeness.

Gaia \citep{gaiadr3} has revolutionized our knowledge of the Solar Neighborhood. This primarily astrometric survey probes deeper and with a higher level of precision than previous efforts from, e.g.~\textit{Hipparcos} \citep{hipparcos}. This has resulted in an exquisite Solar Neighborhood sample that is at least 92\% complete down to spectral-type M9 \citep{GNNS, CNS5}. Such a complete catalog over a larger volume now allows for detailed modeling of \textit{subpopulations} of stars. Larger sample sizes allow to understand the population of stars as a function of parameters like metallicity, age and Galactic population with better precision, but this requires being able to split local \textit{Gaia} stars into these subpopulations. Typically, this requires additional information outside of the astrometric solution (e.g. spectroscopy), which offers a more complete picture of the star (e.g. its metallicity).

Fortunately, multiple spectroscopic surveys have programs that are focusing on the Solar Neighborhood \citep[SDSS-V, DESI;][]{sdssV, DESIMW}. Of interest here are the results coming from SDSS-V's Solar Neighborhood Census (SNC) observing program. The SNC is a dedicated  program that targets stars with both SDSS-V's medium-resolution optical \citep[BOSS;][]{smee13a} and high-resolution near infrared spectrographs \citep[APOGEE;][]{wilson19a} within 100 pc from Sun. The combination of these instruments will provide crucial information on the stellar activity, chemistry, and temperature of the stars within the Solar Neighborhood. Such spectroscopic surveys are not as efficient as a mission like \textit{Gaia} though, so in these current iterations it will not be possible to get a spectrum of all stars within 100 pc. This introduces a fundamental issue with the traditional idea of completeness, which is limited by the least complete type of star in some sample.

With statistical inference methods this is not the case. If we can define the selection function of our spectroscopic survey, i.e.~the probability that a star with some attribute is observed, we can still use the incomplete \textit{Gaia} and SDSS-V samples jointly to model the statistics of subpopulations of stars. There has been much work on selection functions in the advent of \textit{Gaia} \citep[e.g.,][]{rix2021, gaiasf, subsamplegaia} and we can utilize these frameworks to accomplish our goals. The primary goal of this paper is to construct a practical selection-function framework for SDSS-V Solar Neighborhood Census stars relative to the \textit{Gaia} Catalog of Nearby Stars (GCNS), and to show how this framework can be used to infer the distribution of spectroscopically defined stellar subpopulations across the local \textit{Gaia} HR diagram. In this sense, the main scientific contribution of the paper is methodological: we develop a probabilistic framework that links the incomplete SDSS-V spectroscopic sample to the nearly complete \textit{Gaia} 100 pc parent population. The accompanying software implementation is intended as a practical tool for applying this framework, while the scientific examples presented later in the paper are primarily demonstrations of its utility rather than the principal contribution in themselves. \added{Thus, the two main goals of this method are the following.}

\added{First, we seek to determine the probability of observing some subsample of 
stars within SDSS-V. Our problem is further simplified as all of the SNC subsample is within the \textit{Gaia} catalog. 
Multiple studies have discussed how to define a selection function for a sample within a large catalog, where Gaia has been the main motivator. As will be demonstrated in this paper, when using the SNC data the  relevant ``subsample'' is analysis dependent. In the simplest case, it may be defined as the set of SNC targets that have an APOGEE and/or BOSS spectrum in DR19. For other applications, such as H$\alpha$- or metallicity-based
 studies, the relevant subsample is more restrictive: it is the subset 
of observed stars for which the required spectroscopic quantity is 
successfully recovered and passes the adopted quality criteria. So, the 
goal will be to define a generalized method that can take any analysis 
dependent subsample and formulate the selection function.
}

\added{Second, we need some method that translates the stellar parameters to the wider \textit{Gaia} 100 pc sample. What this means is, say we have an observed set of SDSS-V stars with [Fe/H]$< -1$
 and we want to calculate the luminosity function of metal-poor stars. 
To do this, we need to probabilistically determine the likely stars in 
the full \textit{Gaia} 100 pc sample that have [Fe/H]$< -1$.
 This will require the use of the selection function and some model to 
determine this subpopulation probability. With this method though, we 
can fully unlock the power of the SDSS-V dataset such that with just a subset of 
stars with some stellar parameters, we can probabilistically study the 
full population using the \textit{Gaia} dataset.}

The outline of this work is as follows. In \S\ref{sec:data} we provide an overview of SDSS-V targeting in general and of the Solar Neighborhood Census. In \S\ref{sec:method} we describe our methods for determine the selection function and for modeling a subpopulation of stars. In \S\ref{sec:results} we illustrate the utility of our methods through both mock and real examples. These examples are intended to validate and illustrate the framework in practical scientific cases. In \S\ref{sec:improve} we discuss some limitations and room for improvement for the methods. Finally, in \S\ref{sec:summary} we provide a summary and brief conclusions.

\section{Data} \label{sec:data}

\subsection{SDSS-V Targeting Overview}

SDSS-V is an all-sky, multi-epoch spectroscopic survey that will observe over six million objects \citep{sdssV}. It consists of three ``Mappers", which are the top level programs for the survey. The most relevant for our study is the Milky Way Mapper (MWM), which aims to 1) map the stellar populations and chemodynamics
 of the Milky Way to understand its evolution, and 2) probe stellar 
physics and system architecture by observing a variety of stars in the 
Milky Way and Magellanic Clouds. Data for MWM
 is being collected at two observatories; the Sloan Foundation 2.5-m 
telescope at Apache Point Observatory (APO) in New Mexico, USA \citep{APO} and the Irénée du Pont 100-inch telescope at Las Campanas Observatory (LCO) in Chile \citep{LCO}.
 Both observatories are equipped with an optical \citep[BOSS;][]{smee13a} and a near-infrared
\citep[APOGEE;][]{wilson19a} fiber-fed spectrograph. Most importantly, SDSS-V has introduced
 the focal plane system \citep[FPS;][]{FPS}, which utilizes 500 robotic fiber positioners that enable efficient changes between telescope pointings, greatly increasing survey efficiency.

 To plan such a large survey, SDSS-V has opted to algorithmically plan the full survey, such that all observations (consisting of some configuration of the 500 robots; referred to as a ``design") are planned ahead of time. This is accomplished by the survey planning software, \texttt{robostrategy} \citep{blanton2025}. The main inputs to \texttt{robostrategy} are a set of ``cartons", which are collections of targets with a specific scientific goal. Each carton is assigned a value, which helps optimize the number of visits each field receives, and a priority, which sets the order in which targets are assigned to fibers within a given field. Additionally, there are constraints placed on which targets can be assigned to a fiber \citep{medan2025}, which are a function of magnitude, sky location, location in the focal plane, etc.
 
 This context is particularly important when examining any subsample of spectra in SDSS-V. While a single carton may have a well-defined target-selection rule, the final sample available for scientific analysis can differ substantially after the targets pass through \texttt{robostrategy}. In practice, the effective sample may be shaped by at least four distinct stages: carton-level target selection, fiber assignment, actual observation by the survey, and, in some cases, successful recovery of the spectroscopic quantity required for the analysis. These stages should be kept conceptually separate. A source may satisfy the carton criteria but never be assigned to a fiber, it may be assigned but not yet observed in the data release under consideration, or it may have a spectrum but still fail to yield the derived quantity required for a given science case. In this work, we seek to define a selection function that will encapsulate these compounding effects for the Solar Neighborhood Census (SNC).

\subsection{Solar Neighborhood Census}

\begin{figure*}
	\centering
    \includegraphics[width=0.45\textwidth]{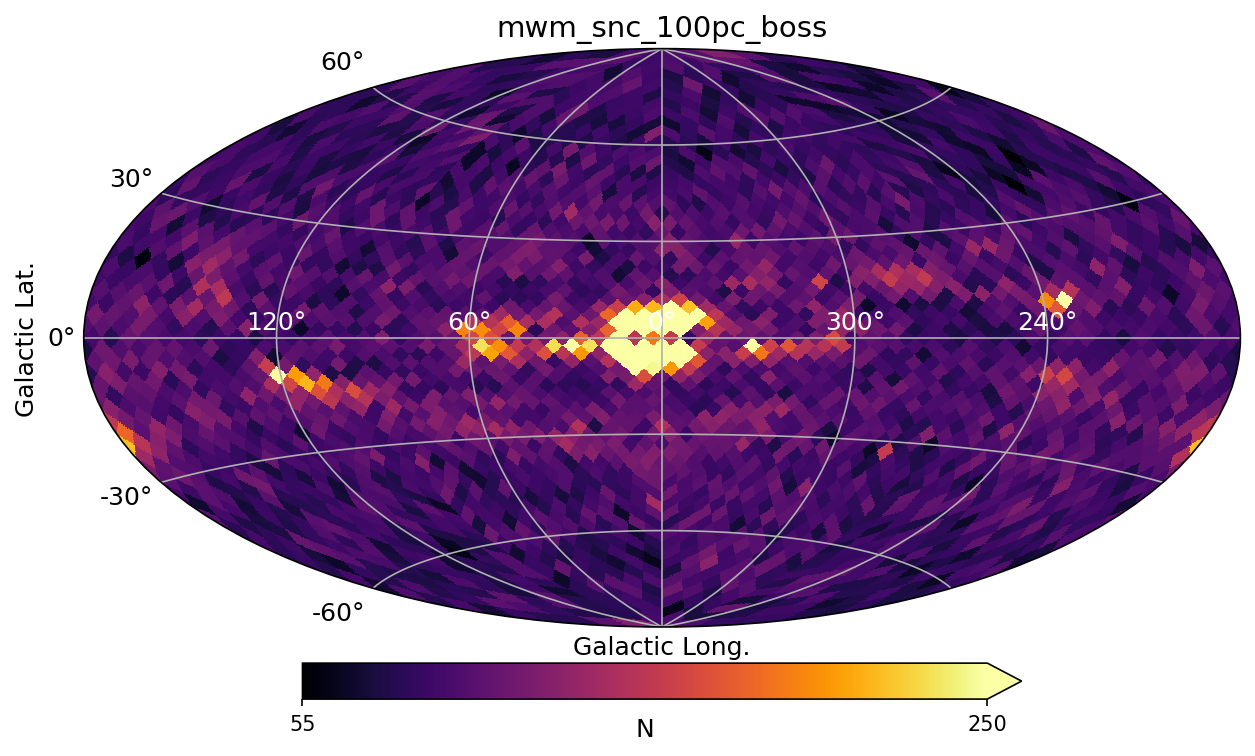}
    \includegraphics[width=0.45\textwidth]{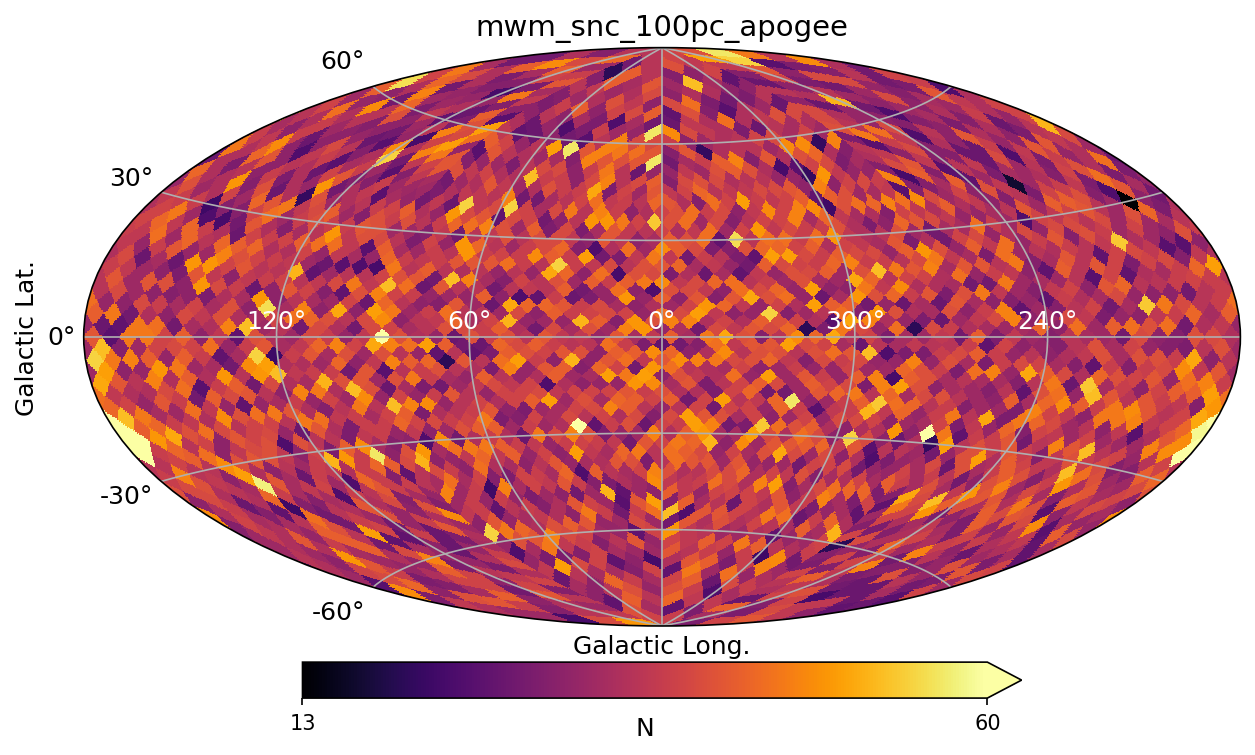}
    \includegraphics[width=0.45\textwidth]{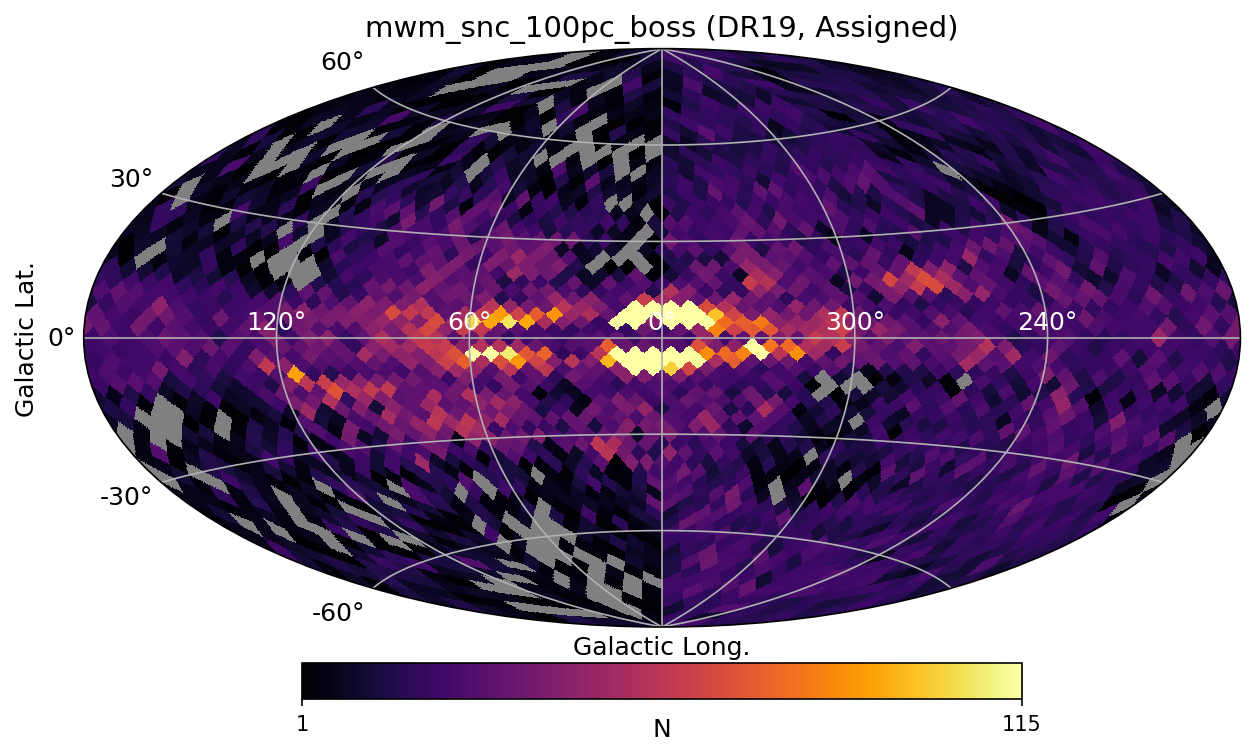}
    \includegraphics[width=0.45\textwidth]{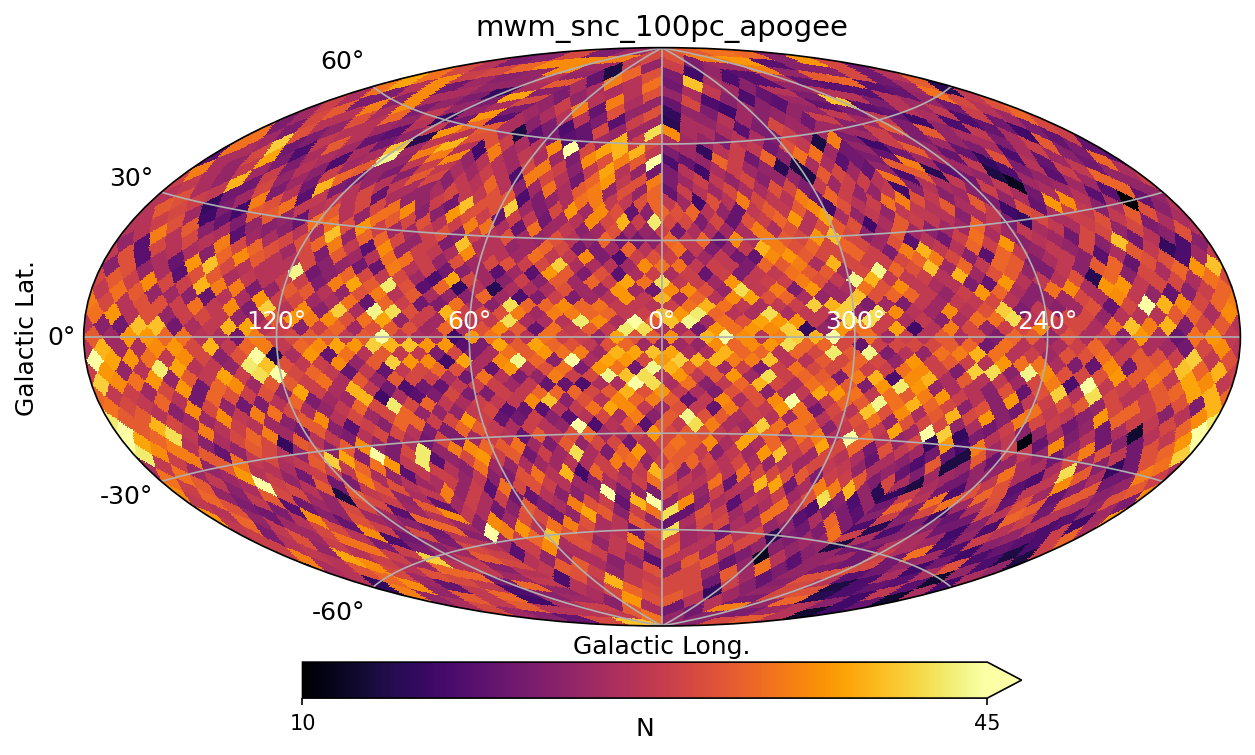}
	\includegraphics[width=0.45\textwidth]{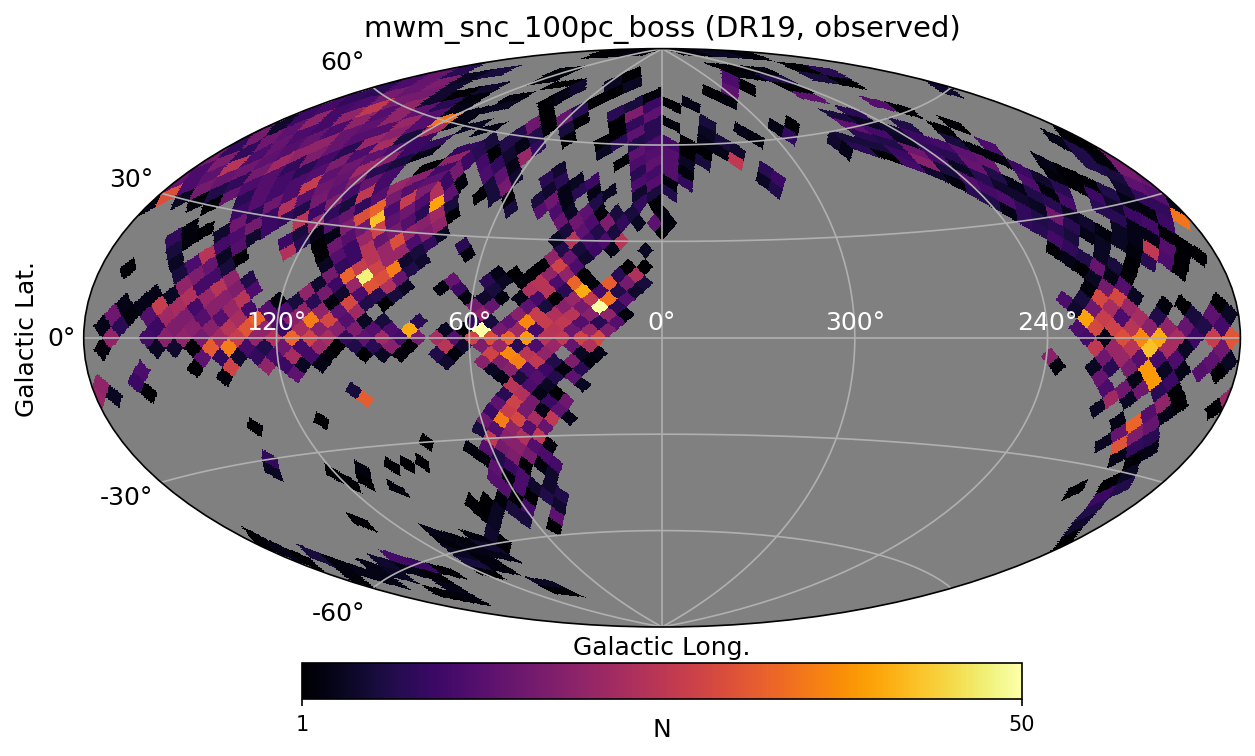}
	\includegraphics[width=0.45\textwidth]{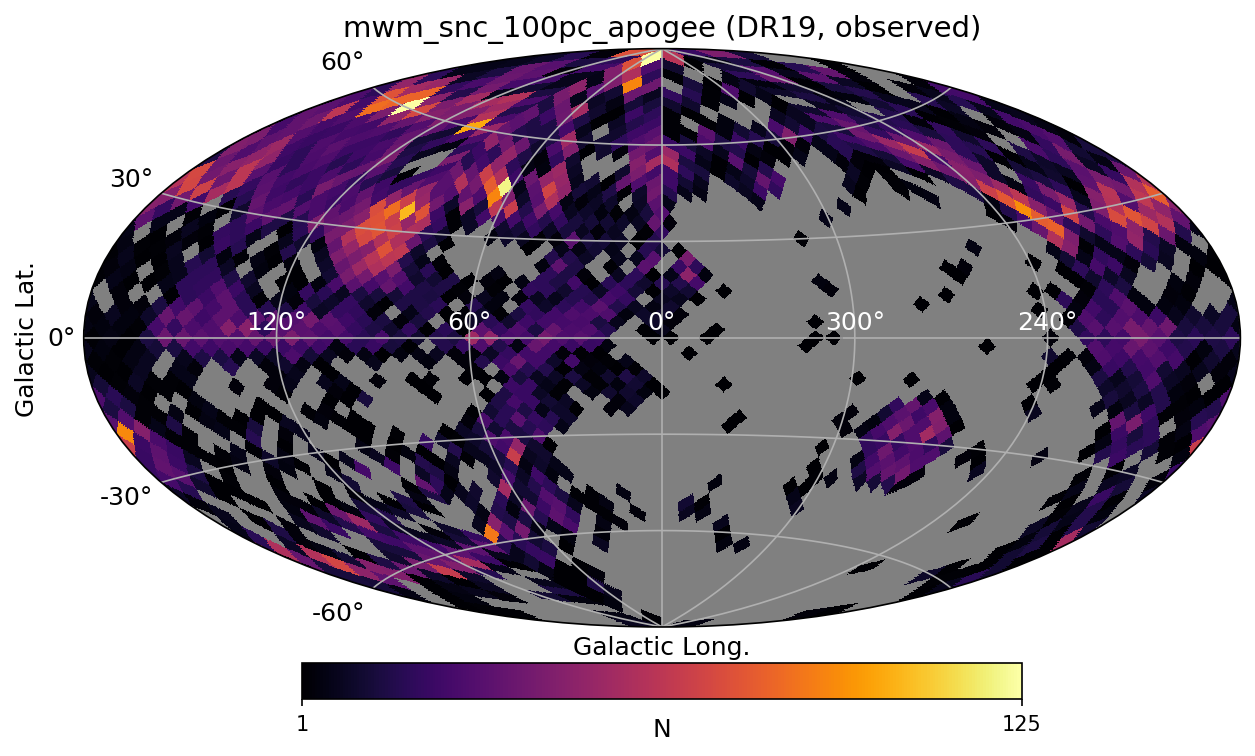}
	\caption{Sky plot of the SDSS-V targets in the \texttt{mwm\_snc\_100pc} cartons for BOSS (left column) and APOGEE (right column). The top row is the sky distribution for the full carton, which serves as the input for \texttt{robostrategy}. The middle row shows the targets that are assigned to a fiber by \texttt{robostrategy} in the DR19 survey plan. The bottom row shows the targets that have a spectrum in DR19.}
	\label{fig:sky_plot}
\end{figure*}

The Solar Neighborhood Census (SNC) seeks to observe a volume limited sample of stars near the Sun. This is broken up into the 100 pc sample and an extension sample, which seeks to observe a larger number of higher mass stars at distances up to 250 pc. In this work, we will be focusing on the 100 pc sample. The carton selection for the SNC 100 pc sample released in DR19 \citep{sdssdr19}, as described in \cite{sdssdr18}, follows:
\begin{enumerate}
    \item $\varpi - \sigma_\varpi > 10$ mas;
    \item \texttt{astrometric\_excess\_noise} $ < 2$ if in one of the following regions:
    \begin{enumerate}
        \item $l \leq 180$ AND $b < -0.139 \ l + 25$ AND $b > 0.139 \ l -25$;
        \item $l > 180$ AND $b > -0.139 \ l + 25$ AND $b < 0.139 \ l -25$;
        \item $\sqrt{(l - 303.2)^2 + 2 \times (b + 44.4) ^ 2} < 5$;
        \item $\sqrt{(l - 280.3)^2 + 2 \times (b + 33.0) ^ 2} < 5$;
    \end{enumerate}
\end{enumerate}
The above is based on the \textit{Gaia} DR2 data, specifically. Future data releases will include target selections based on DR3. The restrictions placed on astrometric noise and Galactic coordinates are meant to put stricter cuts on regions near the Galactic plane, Large and Small Magellanic Clouds. All \textit{Gaia} targets that meet the above cuts are allocated to the carton \texttt{mwm\_snc\_100pc\_boss}. All targets that meet the above \textit{and} have $H < 11$ mag are allocated to the carton \texttt{mwm\_snc\_100pc\_apogee}. The top row of Figure \ref{fig:sky_plot} shows the sky distribution of the resulting cartons used for the DR19 survey plan. For both cartons, there is a relatively uniform distribution of stars across the sky, with the exception of the Galactic center for BOSS. This overdensity will be addressed in the target selection for future data releases.

The middle row of Figure \ref{fig:sky_plot} shows the next stage in the process: the subset of carton-selected targets that are assigned to a fiber in the DR19 survey plan generated by \texttt{robostrategy}. For BOSS, we see that there are many regions where no targets are assigned, particularly in high latitude regions where extragalactic targets have higher priority. Additionally, we see an asymmetry in longitude in the assignments for BOSS, which is due to the higher value of SPectroscopic IDentfication of ERosita Sources (SPIDERS) fields. This assigned sample is therefore already a non-trivial subset of the full carton-level sample.

The bottom row of Figure \ref{fig:sky_plot} shows a third stage: the targets that were not only assigned in the DR19 plan, but that actually have a spectrum in DR19. As DR19 only covers a small portion of the full survey, we are missing large portions of the sky for both APOGEE and BOSS targets. Also, no LCO data is included for BOSS in DR19, leaving a large gap. Finally, the sky is not sampled uniformly over time, which adds to the sparse coverage before the end of the survey. For some science applications, however, even this observed-spectra sample is not yet the final effective sample. A given analysis may additionally require that the spectrum yield a usable derived quantity, such as an H$\alpha$ equivalent width or a reliable stellar metallicity estimate. Thus, the relevant effective selection is not simply the probability that a source is ``included in SDSS-V'', but rather the probability that it passes through the full chain of carton-level target selection, fiber assignment,  observation, and, where required, successful recovery of the spectroscopic quantity used in the analysis. All of this leads to a very complicated selection function for users of the DR19 data. We seek to address this, so the DR19 100 pc data can be used in a statistically robust manner.

\section{Probabilistic Framework}\label{sec:method}

\added{As stated above, SDSS-V's spectra provide  stronger constraints on stellar parameters than the \textit{Gaia} data alone. For our framework, our main goals are to 1) define a flexible, probabilistic framework for describing the selection function, 2) contruct a forward model to determine which \textit{Gaia} 100 pc stars likely belong to various subpopulations within the SNC and 3) use the above results to calculate volume complete statstics. The following sections will describe how we construct and use these methods to meet these goals.}

\subsection{Subsample Selection Function}\label{sec:select_func}

\added{Here} we seek to generally define a subsample selection function that can be analysis dependent. With this is mind, such a selection function can defined as:
\begin{equation}\label{eq:select_func}
    \mathcal{S}_\mathcal{C}^\mathsf{subsample}(q) = \mathcal{S}_\mathcal{C}(q|q \ \mathsf{in \ parent}) \cdot \mathcal{S}_\mathcal{C}^\mathsf{parent}(q)
\end{equation}
where $\mathcal{S}_\mathcal{C}^\mathsf{parent}(q)$ is the probability that a source with  some attributes $q$ will make it into the \textit{Gaia} catalog and $\mathcal{S}_\mathcal{C}(q|q \ \mathsf{in \ parent})$ is the probability that a source will be in the specific analysis subsample within the \textit{Gaia} catalog. $\mathcal{S}_\mathcal{C}^\mathsf{parent}(q)$ has been estimated for $q = \{\alpha, \delta, G\}$ from \citet{gaiasf} and a method for estimating $\mathcal{S}_\mathcal{C}(q|q \ \mathsf{in \ parent})$ has been outlined in \cite{subsamplegaia}. For this work, we will use the precomputed $\mathcal{S}_\mathcal{C}^\mathsf{parent}(q)$ from \cite{gaiasf} as implemented in \texttt{gaiaunlimited}\footnote{\url{https://gaiaunlimited.readthedocs.io/en/latest/}}, where the parent selection function is computed on a HEALPix grid of order 7 as a function of \textit{Gaia} $G$ band magnitude. In general, $>99\%$ of sources have $\mathcal{S}_\mathcal{C}^\mathsf{parent}(q) = 1$, so this is largely negligible, but we include it for completeness. So, the main focus of this work is to adapt the method from \cite{subsamplegaia} to estimate $\mathcal{S}_\mathcal{C}(q|q \ \mathsf{in \ parent})$ for the effective SNC analysis sample under consideration, particularly within 100 pc.

As outlined in \cite{subsamplegaia}, the \added{posterior probability of observing some source within a subsample ($\mathcal{S}_\mathcal{C}(q|q \ \mathsf{in \ parent})$) is best defined by a Beta distribution of the form:}
\begin{equation}\label{eq:beta}
    \mathrm{Beta}(\alpha=k+1, \beta = n-k+1) = \frac{1}{\mathrm{B}(\alpha, \beta)} x^{\alpha - 1} (1-x)^{\beta-1}
\end{equation}
where,
\begin{equation}
    \mathrm{B}(\alpha, \beta) = \int^1_0 t^{\alpha - 1} (1-t)^{\beta-1} dt
\end{equation}
This generally means that the probability of selecting a star in a subsample can directly be calculated from the \textit{Gaia} catalog by counting the total number of objects ($n$) and the number of objects in the subsample ($k$) with some attributes ($q$). For this work, we will define attributes $q = \{\alpha, \delta, G, \added{BP-RP}\}$, i.e.~the Right Ascension, Declination, \textit{Gaia} G band photometry, and \textit{Gaia} $\added{BP-RP}$ color. For the sky location, this is done using HEALPix tessellation. We note that such a HEALPix
 grid does not necessarily align with the fixed grid of fields used by 
SDSS-V. The SDSS-V fields are slightly overlapping and the size of the 
fields change between the two observatories. Because of this, it is 
simpler to model the selection function with a HEALPix grid, though some of the field level decisions may be blended when a HEALPix overlaps between two fields.

\begin{figure*}
	\centering
	\includegraphics[width=\textwidth]{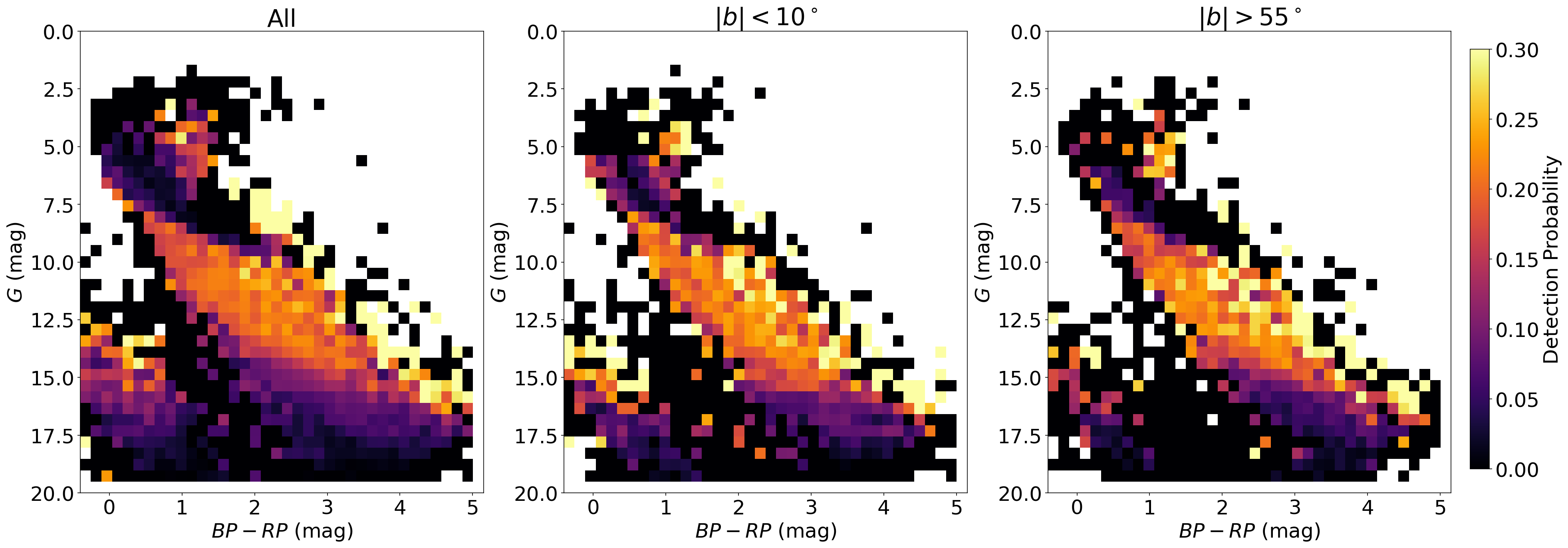}
	\caption{Color-magnitude diagram of the median detection probability, $\mathcal{S}_\mathcal{C}(q|q \ \mathsf{in \ parent})$, for targets with an APOGEE and/or BOSS spectrum from SDSS-V DR19 that are within a \texttt{mwm\_snc\_100pc} carton. The left panel is for targets across the whole sky, the middle panel is for targets with $|b| < 10^\circ$ and the right panel is for targets with $|b| > 55^\circ$.}
	\label{fig:CMD_prob}
\end{figure*}

\added{Additionally, the choice of color is imperative. For this work, we have chosen to demonstrate the method with $BP-RP$, but the accompanying code accommodates $G-RP$ as well. $BP-RP$ has the advantage of avoiding large, overestimation of colors in some cases. For crowded regions or closely separated systems, both the $BP$ and $RP$ aperture photometry will be blended, while the PSF $G$ band photometry will not \citep{Riello2021}. This can cause stars in $G-RP$ to appear significantly to the red of the main sequence, which will place stars in the incorrect region of the HR diagram for the subpopulation methodology to follow. As $BP$ and $RP$ are both blended, the color, while still incorrect, has a shift that is less significant. Because of this we choose to use $BP-RP$ here. We do note that this comes with one major disadvantage. For the lowest-mass M dwarfs, the counts are so, low on the blue end, that the low-counts cause sources to appear bluer than they are. As the majority of our sample will be $G<18$ (Figure \ref{fig:CMD_prob}), this will have a minimal effect on these examples.}

\added{In the above}, we are not concerned with the entirety of the \textit{Gaia} catalog and only aim to find the subsample selection function for 100 pc stars. Fortunately, there is a well defined sample of 100 pc stars in \textit{Gaia} where the overall completeness is \added{fairly} well understood. The \textit{Gaia} Catalog of Nearby Stars \citep[GCNS;][]{GNNS} is a catalog of stars within 100 pc that is 92\% complete down to the hydrogen burning limit. Its completeness can largely be defined by sky varying magnitude limits making it easy to work with when calculating bias-corrected distributions. As a result, our subsample selection function will be relative to the GCNS. This means the star counts from the GCNS will set the total number of objects, $n$, in each bin. For any analysis with the SDSS-V data under consideration (DR19 observed BOSS and/or APOGEE spectra, reliable metallicity measurement, etc.), the specific number of SNC stars considered will set the value of $k$. \added{To enforce that the SDSS-V sample is always a subset of the GCNS, before calculating the selection function all sources not in the GCNS, based on Gaia \texttt{source\_id}, are removed from the user supplied SDSS sample.} This \added{procedure} provides the flexibility to construct a selection function that is analysis dependent. Finally, one change from the method outlined in \cite{subsamplegaia} is that for bins where $k=0$, the probability of selecting a star in SDSS-V will be strictly 0. This is because there are fields that are simply not visited in DR19, so there should strictly be 0 probability of observing such a source.

\added{The above assumes that the incompleteness of the GCNS can largely be modeled in a post-processing step, as $\mathcal{S}_\mathcal{C}^\mathsf{parent}(q)$ \citep{gaiasf} does not account for specific incompleteness regimes in the GCNS. First, due to the magnitude limit of Gaia, the maximum volume probed per spectral type must be accounted for when determining volume complete stellar densities. The method for such an accounting will be discussed in Section \ref{sec:num_dens}. Additionally, due to, largely, astrometric issues at the bright end and for binary systems, the GCNS also misses some hot main sequence stars, stars with high proper motions and binary stars with poor orbital solutions \citep{Fabricius2021, CNS5}. Finally, stars that may have astrometric solutions that confidently place them within 100 pc, but do not have $BP$ and/or $RP$ photometry will be missed by the above. Again, we expect this to most significantly effect the lowest mass sources in the sample. Such effects on the completeness are difficult to model and are not accounted for in this work. This means we expect that there our methods underestimate the incompleteness for the brightest stars ($G\sim 4$ mag), the lowest mass stars (late M dwarfs and L/T dwarfs near the Gaia magnitude limit) and close binaries. For studies that specifically target these regimes, such systemics should be considered when using this method. For the vast majority of the main-sequence though, we expect these methods to perform well and accurately account for the selection effects of the 100 pc sample.}

With this framework, we can examine our detection probability, $\mathcal{S}_\mathcal{C}(q|q \ \mathsf{in \ parent})$. As a baseline sample, we consider all targets with an APOGEE and/or BOSS spectrum from SDSS-V DR19 that are within a \texttt{mwm\_snc\_100pc} carton. Figure \ref{fig:CMD_prob} shows the detection probability across the color-magnitude diagram for the SNC DR19 targets. The subsample selection function is calculated on a HEALPix grid of order 3, bins of \textit{Gaia} G band photometry for $0<G<20$ mag with widths of $0.5$ mag and bins of \textit{Gaia} $\added{BP-RP}$ color for $-0.4 < \added{BP-RP} < \added{5.25}$ mag with bin widths of $\added{0.15}$. The resulting probability of a source being in the subsample is quite consistent with the underlying assignment logic for the survey.

For example, targets with $G< 13$ are not able to get a BOSS spectrum within DR19. This is because they are too bright to be included in observations, as they can contaminate on chip neighbors. In future data releases, these targets will have spectra due to the addition of an offsetting feature \citep{medan2025}. This means these bright targets are only eligible to be assigned to APOGEE. Due to competition with large, APOGEE-led programs like Galactic Genesis, which are predominately in the plane, we would expect a smaller fraction of these bright stars observed in the plane versus the poles. This is consistent with the probabilities in Figure \ref{fig:CMD_prob} when we discriminate by sky location. Similarly, we did not observe any APOGEE targets with $H<7$ in DR19. This is why we see the color-dependent drop in detection probability at the brightest end \added{Additionally, this demonstrates why it is advised to include color as a dimension in the selection function}.

The opposite occurs for the fainter sources, which are primarily targeted with the lower-resolution, BOSS instrument. At high Galactic latitudes, the main competition comes from the extragalactic Black Hole Mapper targets, which have a higher priority for assignment to a fiber. In the plane though, SNC is one of the highest priority optical cartons. This means that out of the plane we are less likely to get an observation because of the competition with BHM, but more likely to observe a star when in the plane. This is well reflected in Figure \ref{fig:CMD_prob}.

The important takeaway from Figure \ref{fig:CMD_prob} is that the complex decisions that inform the survey plan for SDSS-V are captured in the resulting detection probability. This means that these probabilities can be used to account for these complex biases, without having to have a complete understanding of these details. As we will show later, these probabilities can be used to extract vital statistics from the SNC dataset, despite the complex selection that informs it.

\subsection{Subpopulation Probability}\label{sec:subpop_prob}

With our selection function defined, we now want to model a subpopulation of GCNS stars using the observed parameters available from SDSS-V. \added{Here we model the observed SDSS-V SNC stars as an inhomogeneous Poisson point process. In this framework, stars are treated as discrete events whose expected density varies across some space of observables, $q$. This process is described by an intensity function, $\Lambda_{obs}(q|\Theta)$, which gives the expected number density of observed stars at our observable for a set of model parameters, $\Theta$.}

\added{So, the likelihood of observing a set of $N$ stars at locations $q_i$ is:
\begin{equation}
\mathcal{L} =\exp\left(-\int \Lambda_{\rm obs}(q|\Theta)\ dq\right) \prod_{i=1}^{N}\Lambda_{\rm obs}(q_i|\Theta)
\end{equation}
where the product term evaluates the intensity at the location of each observed star, while the exponential term gives the Poisson probability of observing the total number of events implied by the model. Taking the logarithm yields:
\begin{equation}
\log(\mathcal{L}) = \sum_i \log \ \Lambda_{obs}(q_i|\Theta) - \int \Lambda_{obs}(q|\Theta) dq
\end{equation}
Qualitatively, the first term rewards models that place high intensity at the locations of the observed stars. The second term is the expected total number of observed stars predicted by the model and acts as a normalization term that penalizes models that greatly over-predict counts.}

\begin{figure*}
	\centering
	\includegraphics[width=\textwidth]{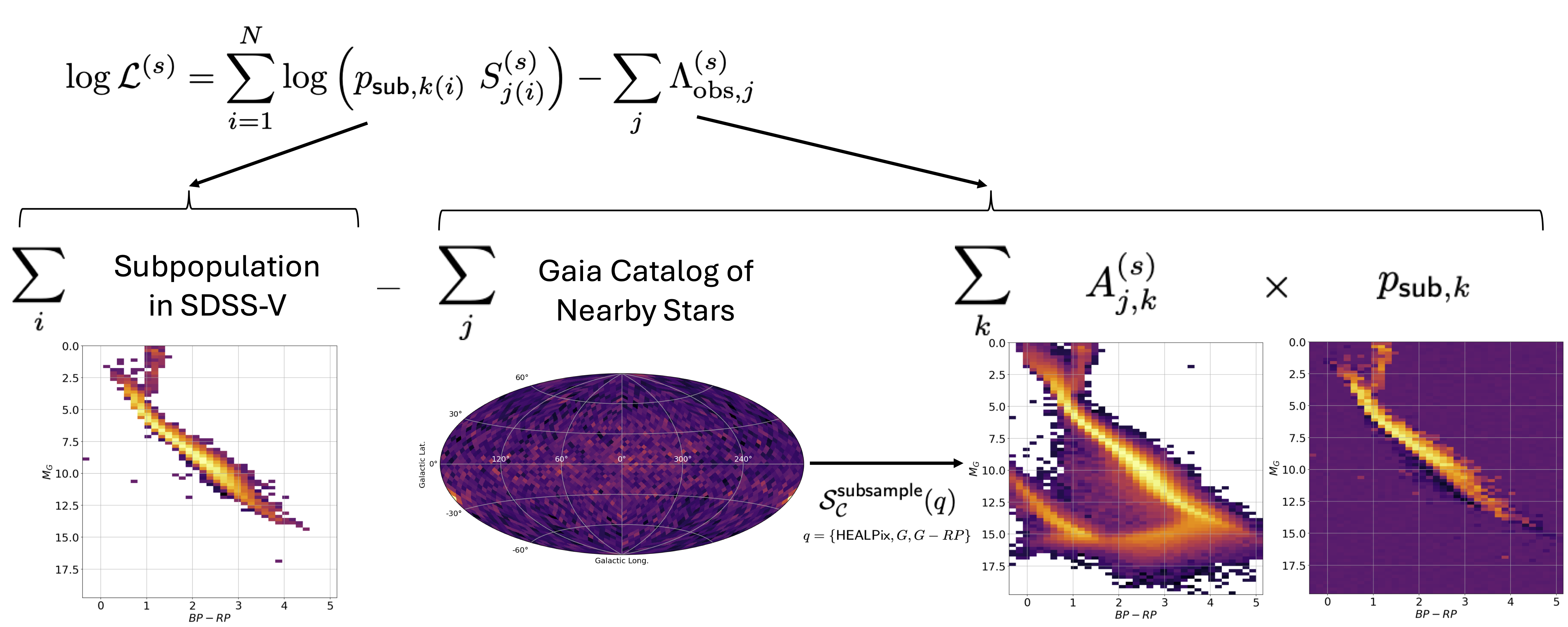}
	\caption{Schematic illustrating the various terms of our subpopulation probability forward model, as described in equations \ref{eq:sel_fact}$-$\ref{eq:loglike}. \added{The first panel shows the HR diagram of the observed SDSS-V SNC sample, where each star is weighted by the subpopulation probability and selection function. The second panel is the sky distribution of the GCNS. The third panel is the HR diagram of the GCNS after applying the selection function. The fourth panel shows the model subpopulation probability across the HR diagram.}}
	\label{fig:model_schematic}
\end{figure*}

\added{Using the above, the goal then is to create a model to predict subpopulations of stars observed within our SDSS-V SNC dataset.} If you have some known selection function ($S$)
 and observations of stars within a subpopulation from SDSS-V, then \added{finding the model that minimizes the above likelihood can be used to} predict the likely distribution of the full subpopulation in a 
modeled space. Here, our model space is going to be the HR diagram, i.e.~$(\added{BP-RP}, \ M_G)$,
 such that we model stellar subpopulations not at the level of 
individual stars, but as a probability field across the HR diagram. This
 choice is motivated by the fact that stellar effective temperature, 
luminosity, mass, metallicity, and evolutionary state generally map in a structured, but non-deterministic, way onto \textit{Gaia} color--magnitude space, both empirically and in stellar-evolution models \citep[e.g.,][]{Bressan2012, Choi2016}. A schematic of our model is shown in Figure \ref{fig:model_schematic} and will be referenced as we describe the various parameters below.

To construct the above likelihood, we will be binning the data in our selection function space ($j$; HEALPix, $G$, $\added{BP-RP}$) and model space ($k$; $\added{BP-RP}$, $M_G$). Additionally, as there is a full posterior for the selection function (eq.~\ref{eq:beta}) and posterior samples of the distance \citep{GNNS} for each object, the likelihood will be calculated for some draw from that posterior, $s$. For the distance posterior from the GCNS, this essentially changes 1) the bin in model space, $k$, the object falls into and 2) if it should be included in likelihood determination at all. For the latter point, if the posterior distance sample is $> 100$ pc, the object is not included in the below calculation.

To calculate the likelihood of our Poisson point process then, we must define the intensity function. To do this, we will define an ``effective selection factor":
\begin{equation}\label{eq:sel_fact}
    A_{j, k}^{(s)} = \sum_{m\in(j, k)} \mathcal{S}_{\mathcal{C}, m}^{\mathsf{subsample}(s)}
\end{equation}
which represents the expected number of stars that SDSS-V would 
observe in bin $(j,k)$ if every GCNS star were subject to the selection 
function. The above is the sum of all subsample selection probabilities, $\mathcal{S}_{\mathcal{C}}^{\mathsf{subsample}}$ (eq.~\ref{eq:select_func}), for some posterior draw for all GCNS stars that are jointly in some model bin \(k\) and selection-function bin \(j\). The sum over all bins would then give the expected number of stars observed in SDSS-V within the full sample under consideration. For example, this could be ``all SNC stars with a reliable ASPCAP measurement". In Figure \ref{fig:model_schematic}, this is demonstrated by taking the full GCNS, weighting each source by the selection probability and then the result is the HR diagram that should match what is observed in the SDSS-V sample.

For the intensity function though, we want to determine the expected number of stars within some \textit{subpopulation} observed in SDSS-V. This can be described as:
\begin{equation}
    \Lambda_{obs, j}^{(s)} = \sum_k A_{j, k}^{(s)} \times p_{\mathsf{sub}, k}
\end{equation}
This equation expresses the key idea of the framework: the observed 
distribution is the \added{product} of the intrinsic subpopulation 
distribution with the survey selection function. Here, we multiply by the probability of a star being in a HR diagram bin ($p_{\mathsf{sub}, k}$), which then reduces the number to the total expected in the subpopulation for that selection function bin. For this method, the values of $p_{\mathsf{sub}, k}$ are what we will want to model and are then free parameters. Figure \ref{fig:model_schematic} shows an example of some grid of subpopulation probabilities, which in this schematic is stars that likely have [Fe/H] $>-0.5$ dex.

So, for a single posterior sample, $s$, each star observed in SDSS-V within that subpopulation \(i\) contributes to the likelihood according to its subsample selection probability and subpopulation probability for the bin it is in. This is shown in the far left plot in Figure \ref{fig:model_schematic}, which shows some observed subpopulation in SDSS-V weighted by $p_{\mathsf{sub}, k} \, \, S_{j}^{(s)}$. Bringing this all together, the likelihood becomes:
\begin{equation}\label{eq:loglike}
\log \mathcal{L}^{(s)} = 
\sum_{i=1}^{N} 
\log \left( p_{\mathsf{sub}, k(i)} \, \, S_{j(i)}^{(s)}\right)
- \sum_j \Lambda_{{\rm obs}, j}^{(s)}
\end{equation}
The goal then is to forward model the values of $p_{\mathsf{sub}, k}$ that maximize this total log-likelihood. These values of $p_{\mathsf{sub}, k}$ will then give us the probability of the subpopulation being in the GCNS across the HR diagram, which will be utilized for future statistical analyses.

To implement the above, we will use MCMC to sample the posterior of the $p_{\mathsf{sub}, k}$ values in our forward model. For a prior on $p_{\mathsf{sub}, k}$, we will assume a beta distribution (eq.~\ref{eq:beta}) with $\alpha=1$ and $\beta$ determined by:
\begin{equation}
    f_o = \mathrm{max}\left( \frac{N_{\mathrm{subpop}}}{N_{\mathrm{data}}}; ~ 0.01 \right)
\end{equation}
\begin{equation}
    \beta = \mathrm{min}\left( \frac{1-f_o}{f_o}; ~ 2 \right)
\end{equation}
Where $N_{\mathrm{subpop}}$ is the number of stars observed by SDSS-V in the subpopulation and $N_{\mathrm{data}}$ is the total number of stars observed by SDSS-V. With the above formulation, this will then result in a beta distribution with mean equal to the observed fraction of the subpopulation, $f_o$. We set a floor on $f_o$ to not have too peaky of a distribution and a minimum of $\beta = 2$ so the result is never a uniform distribution. We choose this prior over a uniform prior, as if a subpopulation is inherently rare (like halo stars, for example), this prior will better weight the subpopulation probability to smaller fractions, especially in regions of very low sampling. If we instead always did a uniform prior, then the resulting mean of 0.5 would greatly over sample stars in low sampled regions.

With this prior, in each draw in the chain we will then resample the posterior for the subsample selection function based on eq.~\ref{eq:beta} and recalculate $M_G$ (which dictates what model bin the object is placed in) based on a random draw from the distance posterior samples from \citet{GNNS} for each object in both the GCNS and the observed subpopulation in SDSS-V. With this, we reconstruct our effective selection factor (eq.~\ref{eq:sel_fact}) and then calculate the log-likelihood (eq.~\ref{eq:loglike}). In practice, the effective selection factor is a very sparse matrix where many bins will equal zero, so we will construct it as a sparse matrix in COOrdinate format. Additionally, as the above can be calculated through a series of matrix operations with quite large arrays, we have implemented our forward model in \texttt{JAX} \citep{jax2018github}, with our MCMC utilizing \texttt{numpyro} \citep{numpyro}. This means automatic differentiation and JIT compilation to GPU / CPU is provided and greatly reduces the runtime of the forward model. For the MCMC, we use The No-U-Turn Sampler \citep[NUTS;][]{NUTS} consisting of two chains each with a warm-up of 5000 draws and then 2000 subsequent samples.

Finally, to better identify the significance of the results, we provide a way to flag subpopulation probability posterior distributions that differ greatly from the prior. We do this through three tests.

First, we perform a KL divergence test \citep{kullback1951information}, defined as:
\begin{equation}
    D_\text{KL}(P \parallel Q) = \sum_{ x \in \mathcal{X} } P(x) \, \log \frac{ P(x) }{ Q(x) }
\end{equation}
Here, the reference distribution, $Q(x)$, is the beta prior and $P(x)$ is the observed posterior. We calculate the above discretely for 50 bins between $[0,1]$. Here, small values of $D_\text{KL}(P \parallel Q)$ signify that $Q(x)$ and $P(x)$ are drawn from the same distribution. To quantify a significant value, we bootstrap a baseline by randomly sampling a beta distribution (same as the prior) for the same number of samples as the posterior and calculating $D_\text{KL}(P \parallel Q)$. If $D_\text{KL}(P \parallel Q) - \overline{D_\text{KL, Baseline}(P \parallel Q)} > 5 \sigma_{D_\text{KL, Baseline}(P \parallel Q)}$, then we consider the posterior distribution to be significantly different than the prior.

We also consider two other statistics; variance and mean shift. For the variance shift, we examine if the variance of the posterior is 0.9 of the expected variance of the beta prior. For the mean shift, we see if we observe $z = |\overline{x_{posterior}} - \overline{x_{prior}}| / \sigma^2_{prior} > 1$. If either the KL divergence, variance shift or mean shift criteria is met, then we flag the posterior distribution as significantly different from the prior. An example of this in shown in the Appendix in Figure \ref{fig:appendix_ex}.

\subsection{Calculating Number Density}\label{sec:num_dens}

One important aspect of working with a volume limited sample is understanding the number density of objects. While the above method provides the probability of selecting a \added{GCNS} target within our volume, it does not define the true maximum volume probed. Indeed, for the low-luminosity objects, regardless of the subsample selection, we will never probe all stars out to 100 pc due to the magnitude limit of Gaia. The maximum volume probed by our sample can be determined using \added{a version of} the classical technique from \citet{Schmidt1968}. For Galactic objects, a further correction of the decrease in the stellar density with increase of distance from the plane must be added \citep{felten1976, tinney1993}, resulting in\added{:
\begin{equation}\label{eq:veff_gen}
    V_{\rm max}=\Omega \int_0^{d_\mathrm{max}} r^2\frac{\rho}{\rho_\odot} dr
\end{equation}
}

\added{A similar technique was used by e.g., \citet{GNNS}, but such implementations do not account for the Sun's height above the Galactic plane. In this case:
\begin{equation}
    \frac{\rho}{\rho_\odot} = \exp\left[ \frac{z_\odot - |z_\odot + r \mathrm{sin}(b)|}{H} \right]
\end{equation}
where $H$ is the thin-disc scale height, $z_\odot$ is the height of the Sun above the Galactic plane and $b$ is the Galactic latitude. With this definition, eq.~\ref{eq:veff_gen} can be solved analytically for a few cases resulting in:
\begin{equation}
V_{\rm max}=\Omega
\left(\frac{H}{k}\right)^3
\begin{cases}
2-\left(\zeta^2+2\zeta+2\right)e^{-\zeta},
&
\mathrm{if \ V_+}
\\[0.8em]
\left(\zeta^2-2\zeta+2\right)e^{\zeta}-2,
&
\mathrm{if \ V_-}
\\[0.8em]
\begin{aligned}
&2e^{\eta}\left(\eta^2+2\right)-2
\\
&\quad
-e^{2\eta-\zeta}
\left(\zeta^2+2\zeta+2\right),
\end{aligned}
&
\mathrm{if \ V_\times}
\end{cases}
\label{eq:veff_zsun}
\end{equation}
where $k = \mathrm{sin}|b|$, $\zeta = \frac{d_\mathrm{max} k}{H}$, $r_{\rm cross} = z_\odot / k$ and $\eta = \frac{r_{\rm cross} k}{H}$. In the above, $\mathrm{V_+}$ applies when $b>0$, $\mathrm{V_-}$ when $b<0$ and $d_{\rm max}\le r_{\rm cross}$, and $\mathrm{V_\times}$ when $b<0$ and $d_{\rm max}> r_{\rm cross}$. In the special case where $b=0$, then in the limit of $k\rightarrow0$, $V_{\rm max}=\Omega \frac{d_\mathrm{max}^3}{3}$.}

\added{For the above, we assume} $H = 365$ pc \citep{GNNS}, \added{$z_\odot = 17$ pc \citep{Karim2017},} $d_\mathrm{max}$ is the maximum distance of detection and $\Omega$ is the area coverage of the survey in steradians. With some magnitude limit,  $G_\mathrm{lim}$, the maximum distance derived is:
\begin{equation}
    d_\mathrm{max} = \mathrm{min}\left( 10^{\frac{G_\mathrm{lim} - M_G}{5} + 1}; 100 \ \mathrm{pc} \right)
\end{equation}
\added{Fortunately, the magnitude limits for the GCNS are well understood, such that we can correct for the above incompleteness issues in a robust manner. To accomplish this, we use} the pre-computed magnitude limits from \cite{GNNS}, specifically at the 80th percentile per healpix bin, \added{which they used in their work to calculate the luminosity function.}

Having this capability in our codebase allows users to define volume complete number densities within bins:
\begin{equation}\label{eq:weight}
    \phi = \sum_\mathrm{bin} \frac{1}{V_\mathrm{max}}
\end{equation}
where, in practice, one would sum over the proportion of stars in a subpopulation. This then would allow, for example, for the luminosity function for stars with [Fe/H]$<-1$ to be determined (see Section \ref{sec:mf}). Like in \cite{GNNS}, we recommend sampling over the full distance posterior of the stars to avoid the Lutz-Kelker bias for a volume-limited sample \citep{lutz1973}. Within our codebase, we include all samples of $V_\mathrm{max}$ such that the number density just needs to be averaged over all samples.

\section{Application to Case Studies}\label{sec:results}

Using the methods described in Section \ref{sec:method}, we can now calculate our subsample selection function, determine subpopulation probabilities across the HR diagram and get volume limited number densities of stars based on SDSS-V observations. In the below sections, we will provide some case studies to validate these methods and demonstrate their utility. Each case study uses its own effective analysis sample, defined according to the quantity being modeled. These examples will show how this work opens the door to future, ground-breaking studies that can be conducted with SDSS-V and \textit{Gaia} data.

\subsection{Case 1: Validation of Method}\label{sec:valid}

\begin{figure*}
	\centering
	\includegraphics[width=0.8\textwidth]{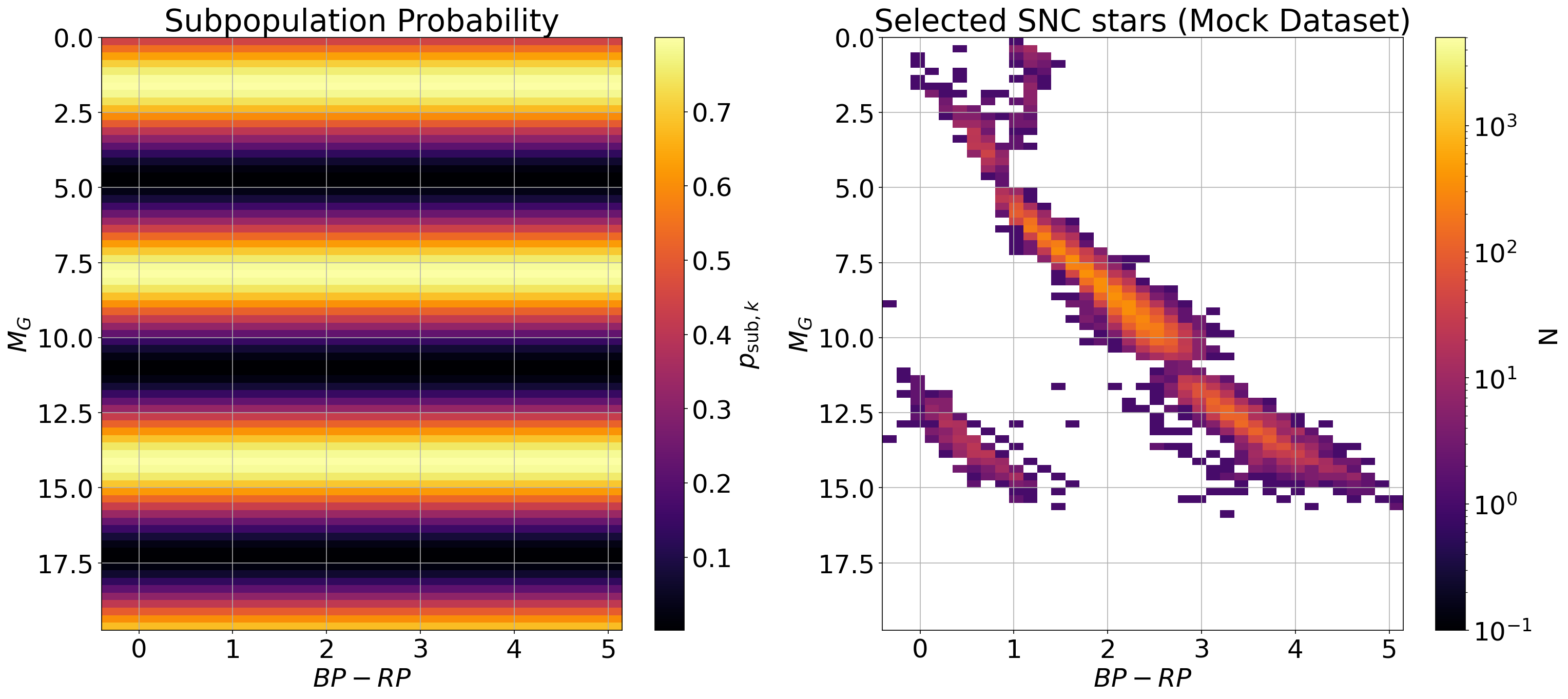}
    \includegraphics[width=0.8\textwidth]{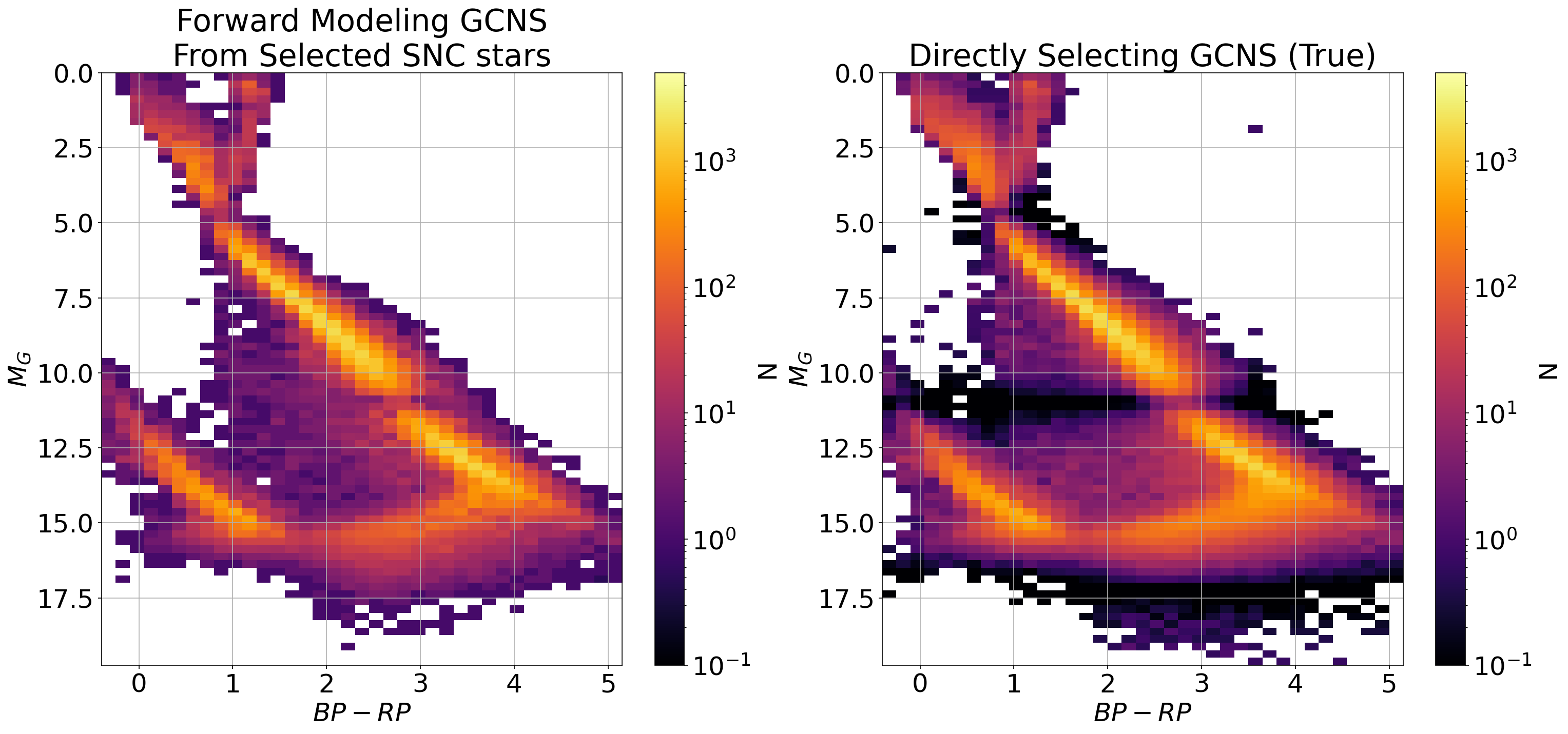}
    \hspace*{-1.1cm}
    \includegraphics[width=0.75\textwidth]{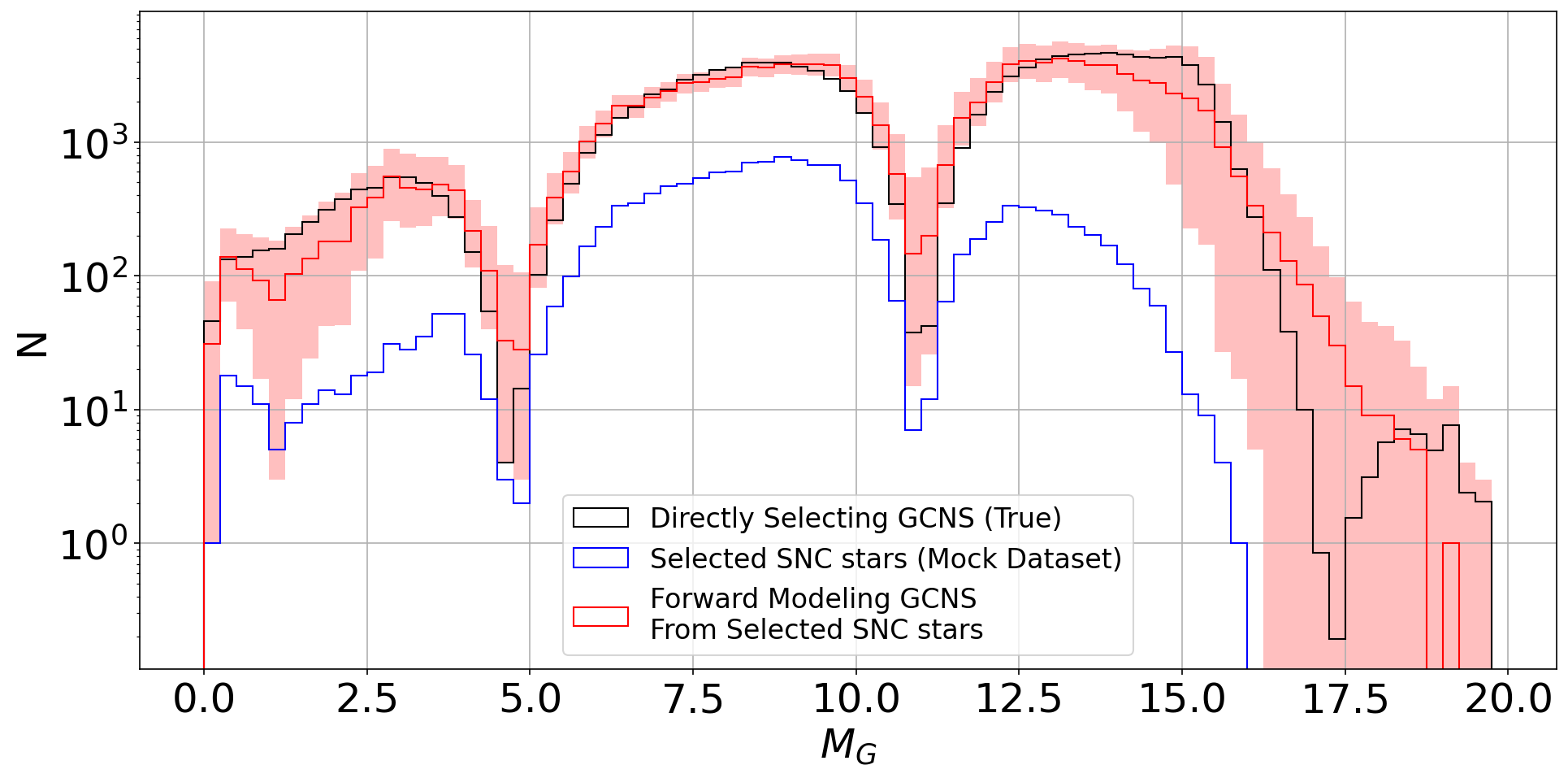}
	\caption{Illustration of the validation of our method with a mock dataset. We model our mock subpopulation as a sine wave with $M_G$. With this, we randomly select stars from the complete GCNS and then our subpopulation observed with SDSS-V is the DR19 \texttt{mwm\_snc\_100pc} stars within that subpopulation (Selected SNC Stars (Mock Dataset)). With our selection function and subpopulation we then forward model the subpopulation probability. Using this forward modeled subpopulation probability, we can select GCNS stars across the HR diagram following a binomial distribution (Forward Modeling GCNS From Selected SNC stars), which should be compared to the true distribution (Directly Selecting GCNS (True)). The bottom panel then shows the 1D luminosity function of all of these samples, where the red shaded region is the 95\% confidence in the forward modeled selection, where we see good agreement with the true distribution.}
	\label{fig:mock_ex}
\end{figure*}

It is necessary to first validate our methods. This can be done by generating a mock subpopulation of stars in SDSS-V. Then, we can see if we are able to recover the true, underlying distribution of the subpopulation in the GCNS. This case will be illustrated through the plots in Figure \ref{fig:mock_ex} and the Python scripts to reproduce this analysis can be accessed on the package's GitHub\footnote{\url{https://github.com/imedan/snc_sf/blob/1.1.1/notebooks/forward_model_GCNS_mock_proof.py}}\added{ or via Zenodo\footnote{\url{https://doi.org/10.5281/zenodo.21285596}} \citep{snc_sf_zenodo}}.

First, we define an arbitrary subpopulation probability across the HR diagram. Here, we model it as a sine wave with $M_G$. This is of course not physical, but demonstrates how our method can recover arbitrary distributions. This mock subpopulation probability is shown in the top left panel of Figure \ref{fig:mock_ex}. With this, we randomly select stars from the complete GCNS by weighting each object by its mock subpopulation probability. Our mock dataset is then the SDSS-V stars in the DR19 \texttt{mwm\_snc\_100pc} cartons that are within this mock GCNS subpopulation. This mock dataset is shown in the top right panel of Figure \ref{fig:mock_ex}.

Next we calculate the selection function and forward model the subpopulation probability across the HR diagram. The selection function is calculated based on the GCNS and the full DR19 \texttt{mwm\_snc\_100pc} dataset, as described in Section \ref{sec:select_func}. The selection function is \added{calculated} in this manner, as the SDSS-V 100 pc dataset is what the subpopulation is being drawn from. Then, we forward model the subpopulation probability based on the observations in our mock dataset, as described in Section \ref{sec:subpop_prob}. With the forward modeled subpopulation probability posterior samples, we can then randomly draw from the full GCNS based on sampling from a binomial distribution. The median of these draws is shown in the left plot in the middle row of Figure \ref{fig:mock_ex}. For comparison, the true HR diagram distribution expected for this subpopulation is shown in the right plot in the middle row of Figure \ref{fig:mock_ex}. Here, we see that the forward model generally recovers the distributions, except in the low number regions where we see the forward model overestimates the number of stars, on average.

This is only the median result though. Indeed, we have the full variance in the estimate across the HR diagram. This is difficult to visualize in two-dimensions, so the bottom panel of Figure \ref{fig:mock_ex} shows the 1D luminosity function of the sample. Here, we now plot the 95\% confidence region (red shaded region) to show the variance on our forward model result. With this, we see that the forward model correctly recovers the true, underlying distribution consistently when the errors are considered. This demonstrates that this forward modeling procedure, which depends on the selection function, allows us to accurately determine the distributions of arbitrary subpopulations of the GCNS based on observations from SDSS-V. We do note that in low-count regions, the results will remain prior dominated.

\subsection{Case 2: H$\alpha$ Presence Across the HR Diagram}\label{sec:halpha}

\begin{figure*}
	\centering

    \includegraphics[width=\linewidth]{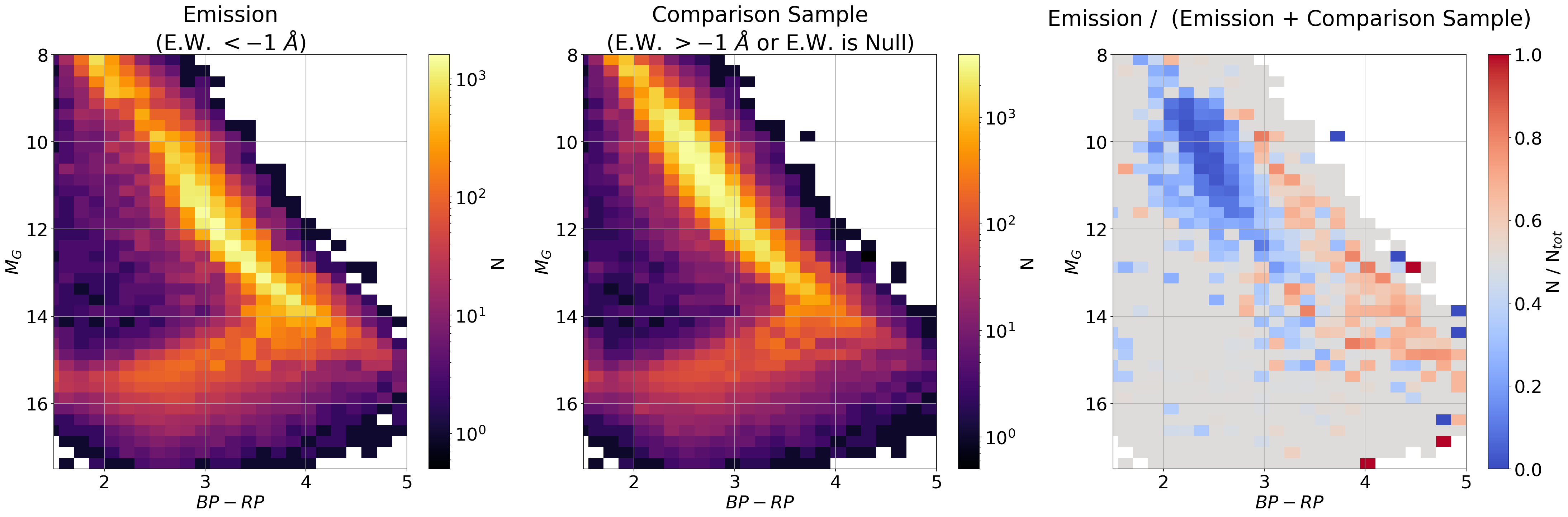}
	\caption{HR diagrams for stars with H$\alpha$ emission (E.W. $< -1$ \ \AA; left panel), or H$\alpha$ absorption or no usable H$\alpha$ measurement  (E.W. $> -1$ \ \AA or E.W. is Null; middle panel). The HR diagrams are the 50th percentile of star counts for the emission and comparison sample after randomly sampling from the GCNS using the respective subpopulation probabilities. The right panel shows the ratio of the emission sample to the total sample, demonstrating that H$\alpha$ emission stars dominate at the lowest masses.
    }
	\label{fig:halpha_ex}
\end{figure*}

Now that the method has been validated, we will demonstrate a more physically motivated example. Here we will examine subpopulations of stars with different H$\alpha$ properties across the HR diagram. We use the SDSS DR19 LineForest \citep{lineforest}, which is a catalog that provides equivalent widths (E.W.) for a variety of lines in SDSS-V optical spectra. In LineForest, negative equivalent widths indicate emission, while positive values indicate absorption. For this application, the relevant effective analysis sample that the selection function is based on is not simply the set of stars in the SNC carton, but the subset of \texttt{mwm\_snc\_100pc} stars that have a BOSS spectrum in DR19 and are processed by LineForest. This case will be illustrated through the plots in Figure \ref{fig:halpha_ex} and the Python script to reproduce this analysis can be accessed on the package's GitHub\footnote{\url{https://github.com/imedan/snc_sf/blob/1.1.1/notebooks/forward_model_halpha_ex.py}}\added{ or via Zenodo\footnote{\url{https://doi.org/10.5281/zenodo.21285596}} \citep{snc_sf_zenodo}}.

For this example, we divide the sample into two subpopulations: H$\alpha$ emission sources (E.W. $< -1 \ \AA$) and a comparison sample defined by E.W. $> -1 \ \AA$ or no usable H$\alpha$ (e.g.~E.W.~is null in LineForest). We note that this second category should be interpreted with care: it combines stars with measured H$\alpha$ absorption and stars for which the relevant quantity is not available in the catalog. In that sense, it is best regarded as an operational comparison sample rather than as a fully homogeneous physical subpopulation. With these definitions, we then forward model our subpopulation probabilities and randomly select stars from the GCNS, as was done in Section \ref{sec:valid}. The results for the H$\alpha$ emission sources and the comparison sample are shown in Figure \ref{fig:halpha_ex}.

Specifically, we focus on the low-mass stars, as we expect a much larger portion of low-mass stars to be active, relative to higher mass stars \citep{lineforest}. The left and middle panel of Figure \ref{fig:halpha_ex} shows the 50th percentile of star counts for the emission and comparison sample after randomly sampling from the GCNS using the subpopulation probabilities. In these HR diagrams, we clearly see that there is a change in the luminosity distribution depending on if H$\alpha$ is observed in emission or not. For the comparison sample, the peak is at $M_G\sim 11$ mag, which we expect for the general field population \citep{GNNS}. For the H$\alpha$ emission stars though, this peak shifts to $M_G\sim 12$ mag, which is significantly fainter. To see this more clearly, the right panel of Figure \ref{fig:halpha_ex} shows the ratio of the emission sample to the total sample. Additionally, we see here that H$\alpha$ emission stars appear redder/overluminous compared to the comparison sample and dominate at the lowest masses.

\subsection{Case 3: Change in \added{Stellar Density} with Metallicity}\label{sec:mf}

This case is likely the most astrophysically relevant example presented here and is intended to illustrate a key science application of the SNC: examining how the \added{relative stellar density} of the low-mass \added{stars varries} with metallicity. With the framework developed above, this becomes possible with the DR19 data set, provided that one can define a clean sample of metallicity measurements and estimate stellar masses for the corresponding stars.  \added{A more significant analysis would be to then relate these measurements to the IMF slope at different metallicities. Such work is out of scope of this paper, but is a planned, key scientific analysis of the SNC.} As a note, the Python script to reproduce this analysis can be accessed on the package's GitHub\footnote{\url{https://github.com/imedan/snc_sf/blob/1.1.1/notebooks/forward_model_MF.py}}\added{ or via Zenodo\footnote{\url{https://doi.org/10.5281/zenodo.21285596}} \citep{snc_sf_zenodo}.}

For the metallicities, we will be using the DR19 data, specifically those compiled in the \texttt{astraMWMLite} file\footnote{\url{https://www.sdss.org/dr19/mwm/astra/pipelines-in-astra/bestparams/}} \citep{sdssdr19}. This file uses a logical set of preferences that take stellar parameter estimates from different pipelines and aggregates them into a file to have one set of stellar parameters for each target. In this case, the relevant effective analysis sample is more restrictive than the general SNC spectroscopic sample: a star must belong to the SNC parent sample, be assigned and observed in DR19, and also yield a usable metallicity estimate in the adopted parameter pipeline. Accordingly, the effective selection function for this science case includes the additional layer of successful recovery of the metallicity quantity used in the analysis. Particularly, we put the following constraints on the stars in the file:
\begin{equation}
\begin{split}
    &\mathrm{slam} \in pipeline, \\
    &\mathrm{aspcap} \in pipeline,\\
    &T_{\rm eff} > 3200 \ K
\end{split}
\end{equation}
The above selects stars from the SLAM \citep{slam} and ASPCAP pipelines, which have the best stellar parameters for K/M dwarfs. It has been shown that ASPCAP metallicities for M dwarfs are too metal-poor \citep{slam}, so these cannot be directly combined as is. To correct this trend, we use a sample of wide binary pairs from \citet{elbadry2021} \added{to measure difference in metallicity between a higher mass primary and an M dwarf secondary (see Appendix \ref{app:metal_corr}). This results in the correction:}
\begin{equation}
\begin{split}
\delta [\mathrm{Fe/H}] ={}& -6.095 \times 10^{-11} \, T_{\mathrm{eff}}^3
+ 9.164 \times 10^{-7} \, T_{\mathrm{eff}}^2 \\
& - 4.564 \times 10^{-3} \, T_{\mathrm{eff}} + 7.538
\end{split}
\end{equation}
We use the above to correct the ASPCAP metallicities for $T_{\rm eff} < 4600$ K, so we can use the SLAM and ASPCAP data together.

We also apply a prefiltering stage for this example. That is, we prefilter both the GCNS and, correspondingly, the observed analysis sample before calculating the selection function and modeling the subsample selection probabilities. In this case, we make \added{an} attempt to remove binaries by removing stars with $\texttt{ruwe} \geq 1.4$, $\texttt{ipd\_frac\_multi\_peak} \geq 0$ and stars that appear overluminous according to \cite{way2026}. Finally, we remove likely white dwarfs from the sample by removing stars with $M_G \geq 10 \times (G - RP) + 5$. With this filtering, the parent sample and the metallicity analysis sample are defined consistently, and we can then perform the selection-function and subpopulation-probability procedure, as described above, in bins of metallicity.

With these subpopulation probabilities, we will now calculate the \added{stellar density as a function of mass and metallicity}. This involves first estimating the mass of the objects in the GCNS. For this, we use the 2MASS absolute $K_s$-band relation from \citet{mann2019}, as this will be reliable for $< 0.7 \ M_\odot$. However, not all stars in the GCNS have a 2MASS measurement so we calculate all masses for those stars with 2MASS photometry, and use these results to linearly interpolate the mass of the others stars in the ${M_G, M_{BP}, M_{RP}}$ space. Using this interpolation we can predict a mass for every K- and M-dwarf in the GCNS and calculate the mass function at different metallicity. This is done through the following procedure.

Similar to the above, for each posterior sample of the subpopulation probability, we randomly draw from the full GCNS based on sampling from a binomial distribution. We then bin the stars in mass and account for the effective volume as described in Section \ref{sec:num_dens}. We do this for all draws from the distance posterior included in the GCNS. At each iteration \added{we then calculate the maximum volume corrected number density of stars.}

\begin{figure*}
	\centering
	\includegraphics[width=\linewidth]{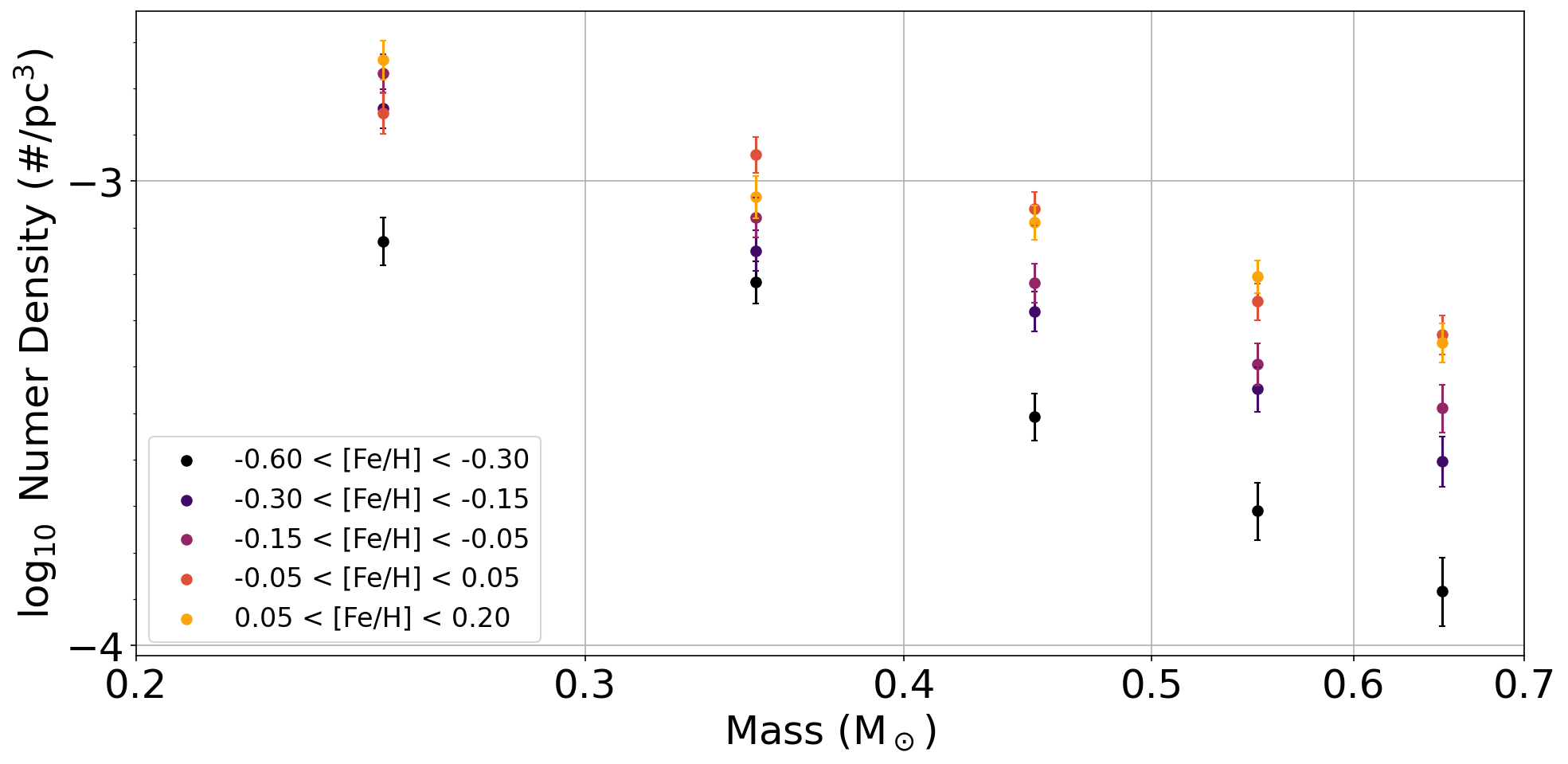}
	\caption{\added{Stellar number density as a function of mass} for different metallicity bins as determined from sampling from the subpopulation probabilities informed by SLAM and ASPCAP data from SDSS-V DR19. The errorbars show the 95th percentile confidence region for these results.}
	\label{fig:mass_ex}
\end{figure*}

Figure \ref{fig:mass_ex} shows \added{the resulting stellar number density as a function of mass for different metallicity bins}. As \added{expected}, we see that the number density of stars increases with increasing metallicity across the entire mass range. Additionally, we find for all metallicity bins that number density increases for lower masses. What is most interesting though is we observe \added{an apparent} change in the slope of the mass function with metallicity. \added{This appears similar to the trend observed in \citet{li2023}, where they found that} the lower metallicity bins are bottom-light (i.e.~slope is shallower than the \citet{kroupa2001} IMF) and stars just below solar metallicity are more bottom-heavy (i.e.~slope is steeper than the \citet{kroupa2001} IMF).

\added{We note that while these results qualitatively match these trends, they differ in one significant aspect. Here, we are only showing the stellar number densities for nominally single field stars, as our prefilitering step specifically removes likely binaries from the sample. To make a direct comparison, this analysis needs to account for the companion fraction as a function of \textit{both} mass and metallicity. Additionally, we would need to incorporate a forward model of the IMF in this analysis to well determine the slopes, and possible mass breaks, of these relations.}

\added{Additionally, the metallicities used here may not be sufficiently accurate for such an analysis.} In this case \added{they are likely insufficient} given the calibrations that had to be done to ASPCAP
 and the limited valid range of the SLAM. Better parameters of low-mass 
stars from SDSS-V data are needed not only to eliminate such systematic 
errors in this regime, but also to allow this method to be applied at 
even lower metallicities.

\added{In the future, with improved metallicities of low-mass stars over a wider range, such a detailed IMF analysis is possible with the use of the method outlined in this paper. This example clearly shows such trends are apparent in the data. With a more robust accounting of binaries and an IMF forward model,} this framework will provide the tools to determine such mass functions 
and to map this trend over a wide range of metallic abundance and 
Galactic environments.

\section{Improvements to Method}\label{sec:improve}

While the method developed in this work is capable of producing statistically robust results in a variety of examples, there are several caveats and possible improvements that should be considered in future applications. These limitations do not invalidate the framework, but they do define the conditions under which it should be interpreted and the directions in which it can be extended.

A first major issue is the role of unresolved binaries and higher-order multiple systems. The method relies on the assumption that position in the HR diagram can be used as a probabilistic predictor of membership in a spectroscopically defined subpopulation. This assumption is only approximate when binaries are present. Unresolved multiple systems can appear systematically overluminous and sometimes redder than comparable single stars, shifting them into regions of the HR diagram that may be associated with different stellar parameters, ages, or metallicities. This is especially problematic for science cases based on metallicity, where unresolved binaries may be mapped preferentially onto more metal-rich single-star loci and thereby distort the inferred subpopulation probabilities. For this reason, it is often advisable to remove likely binaries, or at least to model them explicitly, as was attempted in the example in Section \ref{sec:mf}. A natural improvement for future work would be to include a dedicated binary component in the forward model, so that single stars and unresolved multiples are treated as distinct latent populations rather than forcing both into the same HR-diagram mapping.

A further limitation concerns the assumption that position on the HR 
diagram serves as a unique and stable proxy for the stellar parameters 
inferred from SDSS-V spectra. While this mapping is generally reliable 
for single main-sequence stars, it breaks down in regions of the 
color-magnitude diagram where stellar populations with distinct physical
 properties overlap. For instance, young pre-main-sequence stars and older, metal-rich main-sequence stars can occupy similar regions despite having very different 
ages, activity levels, and chemical compositions. In such cases, the 
subpopulation probability will conflate physically distinct groups of 
stars, potentially biasing the inferred population statistics. Future 
work should explore augmenting the model space beyond the 
two-dimensional HR diagram, for example by incorporating additional 
photometric information such as near-infrared colors or variability 
metrics available from Gaia \citep{gaiadr3},
 to better disentangle overlapping populations. Users of this framework 
should exercise caution when interpreting subpopulation probabilities in
 such degenerate regions of the HR diagram.

The current framework is also sensitive to the adopted binning scheme. The selection function and the subpopulation probabilities are estimated on a discrete grid in sky position, magnitude, and color, and the resulting inferences can depend on the adopted HEALPix order and the bin widths in $G$ and $\added{BP-RP}$. If the bins are too coarse, real structure may be washed out; if they are too fine, the inference becomes noisy and dominated by small-number statistics. This trade-off is especially important in underpopulated regions of the HR diagram or in science cases involving rare subpopulations. Some of this is mitigated by fully sampling the posterior of the selection function distribution, but future use of this method should include explicit sensitivity tests to the adopted binning scheme to ensure it is appropriate for the analysis at hand.

On the \textit{Gaia} side, the method relies on the photometry and astrometry of GCNS stars being sufficiently precise relative to the bin sizes used in the model. If this is not the case, or if there is blending from nearby sources, stars may be scattered into nonphysical regions of the HR diagram. This effect is particularly relevant for faint sources, crowded regions, and stars with problematic photometric solutions. In addition, the current framework implicitly assumes that the parent sample is characterized well enough that these effects are subdominant. A clear improvement would be to propagate observational uncertainties more explicitly, for example by convolving the HR-diagram model with the measurement uncertainties in photometry and color, rather than treating each star as occupying a single exact location. For the examples in this paper, these errors are typically much smaller than the bin sizes. But such factors should be considered when using this method for future analyses.

A further limitation is related to the distinction between lack of survey coverage and lack of successful detections. In the current framework, both situations may enter through low or vanishing counts in a given bin, but they are not physically equivalent. A bin may be empty because that region of the sky or HR diagram was never observed in DR19, or because sources in that region were observed but did not yield the quantity required for the analysis. These two cases should ideally be tracked separately, especially in sparse parts of parameter space. A future improvement would be to propagate explicit masks for survey footprint and observation status, so that measurement incompleteness is not conflated with purely geometric or scheduling incompleteness.

Finally, the method could be extended in a more ambitious way by replacing the current bin-based formulation with a hierarchical generative model for the joint photometric and spectroscopic data. Such a model could simultaneously describe the parent GCNS population, the layered SDSS-V selection process, the measurement-success probability for each pipeline product, and the latent astrophysical subpopulations of interest. We emphasize that the present framework is a practical and already useful approximation to this broader program though, and still serves as an important tool when working with the SNC dataset. A future version should aim to reduce dependence on hard cuts, ad hoc binning, and analysis-specific bookkeeping, while preserving the interpretability and computational tractability that make the current method attractive.

In summary, the main areas for improvement are: (i) a more explicit treatment of binaries and multiples; (ii) a more flexible description of subpopulation probability beyond the HR diagram alone; (iii) sensitivity tests with respect to priors and binning choices; (iv) a more robust treatment of photometric uncertainties, blending, and outliers in the GCNS; (v) a more explicit distinction between lack of sky coverage and reliable measurement; and (vi) a move to a full hierarchical generative model. Addressing these points would substantially strengthen the framework, but require substantial work outside the current scope of this study. These issues and areas of improvement should be considered when using the method in its current form and interpreting any scientific results.

Despite these areas for improvement, we still want to emphasize that the current method serves as a practical and extremely useful analysis tool for the SDSS-V SNC data. The above results demonstrate its clear utility and the types of analyses that can be conducted with the method in its current form.


\section{Summary and Future Work}\label{sec:summary}

In this work, we examine the sample of stars in the Solar Neighborhood ($d < 100$ pc) being observed with SDSS-V. We demonstrate that due to the many competing programs that make up the entirety of the survey, along with the constraints placed on fiber assignments, there are severe selection effects in the resulting observed sample that need to be accounted for before using these data to perform statistical studies. The central goal of this paper is therefore to define a practical selection-function framework for the SNC relative to the GCNS, so that the incomplete SDSS-V spectroscopic sample can be used in a statistically rigorous way together with the nearly complete \textit{Gaia} 100 pc census.

To accomplish this, we use the method from \cite{subsamplegaia} to determine the probability of a Gaia source being observed in SDSS-V. The main adaptation in the method for that, instead of determining this relative to the full Gaia catalog, we use the GCNS as the base sample that our stars are drawn from. We demonstrate that with this, the resulting selection function well matches what we expect from the survey planning constraints and competing programs. This selection-function formalism constitutes the first main contribution of the paper, because it provides the statistical bridge between the observed SDSS-V targets and the parent Solar Neighborhood population.

In addition, we define a new parameter: a subpopulation probability. This is defined as a probability across the HR diagram that a GCNS member belongs to some SDSS-V defined subpopulation  (e.g.~$[\mathrm{Fe/H}] < -1$). When this is combined with the above selection function, we can use a Poisson point process to forward model the likely observed stars in SDSS-V from that subpopulation in the GCNS. The result is a discrete grid of probabilities across the HR diagram, where each bin has a full posterior estimate for this subpopulation probability that can then be used to sample from the GCNS and perform a variety of analyses. Taken together, the selection-function formalism and this subpopulation-probability framework form the main scientific contribution of this work: a probabilistic method for inferring the distribution of spectroscopically defined stellar subpopulations across the local \textit{Gaia} HR diagram despite the incompleteness of the spectroscopic survey.

In this paper, we use this method to demonstrate a few applications of the method. This includes a proof with a mock dataset, looking at the difference in luminosity distribution for H$\alpha$ emitters vs.~non-emitters, and, most interestingly, how the \added{stellar number density} at the low-mass end varies with metallicity. For this final example, \added{this example serves as an important starting point for analysis of the IMF and how it changes with metallicity}. These case studies are included primarily to demonstrate the flexibility and scientific usefulness of the framework, rather than to constitute the principal contribution of the paper.

Finally, an important practical outcome of this work is the public software implementation associated with the method\footnote{\url{https://github.com/imedan/snc_sf}; \url{https://doi.org/10.5281/zenodo.21285596}}\added{ \citep{snc_sf_zenodo}}. The code base released with this paper should not be viewed merely as an auxiliary product, but as a concrete implementation of the statistical framework developed here for the SDSS-V Solar Neighborhood Census. In that sense, the software is an integral part of the reproducibility and broader scientific utility of the methodology.

The package is designed to support the main steps of the analysis described in this paper: construction of the observed SNC sample from SDSS-V products, matching to the GCNS parent catalog, estimation of the effective selection function, and forward modeling of spectroscopically defined subpopulations across the local \textit{Gaia} HR diagram. The example workflow included in Appendix \ref{app:example} is intended to demonstrate that the framework can be applied directly to real DR19 data products with the released code. Additionally, included in the full repository released with this paper are Python script that replicate the full analyses shown in Section \ref{sec:results}.

From a computational point of view, the software reflects a deliberate effort to make the forward model tractable at realistic sample sizes. In particular, the effective selection factor is represented as a sparse matrix, and the likelihood evaluation is implemented through matrix operations in \texttt{JAX}, with posterior sampling performed via \texttt{numpyro}. This provides automatic differentiation, just-in-time compilation, and efficient execution on available CPU or GPU hardware. These choices are important because they make the probabilistic framework practical to use rather than merely formal.

At the same time, there are multiple improvements that could be made to the method to strengthen the long-term impact of the software package (see Section \ref{sec:improve}). These are planned to be active areas of development as we grow the framework with future data releases of SDSS-V. So, the software should be understood as extensible rather than final. The code can be expanded to incorporate better handling of uncertainties, more realistic models of binaries and multiples, and richer definitions of subpopulation probability beyond the HR diagram alone. The present release therefore serves both as a reproducible implementation of the current framework and as a foundation for future methodological development.

Overall, the software released with this work substantially enhances the scientific value of the paper. It translates the statistical framework into a practical research tool, supports reproducibility, and provides the community with a critical tool that can now be used for future studies of the Solar Neighborhood based on SDSS-V and \textit{Gaia}.

\begin{acknowledgments}

J.G.F-T gratefully acknowledges the support provided by ANID Fondecyt Regular No. 1260371.

R. L-V. acknowledges support from Secretar\'ia de Ciencia, Humanidades, Tecnolog\'ia e Inovacci\'on (SECIHTI) through a postdoctoral fellowship within the program ``Estancias posdoctorales por M\'exico''.

Funding for the Sloan Digital Sky Survey V has been provided by the Alfred P. Sloan Foundation, the Heising-Simons Foundation, the National Science Foundation, and the Participating Institutions. SDSS acknowledges support and resources from the Center for High-Performance Computing at the University of Utah. SDSS telescopes are located at Apache Point Observatory, funded by the Astrophysical Research Consortium and operated by New Mexico State University, and at Las Campanas Observatory, operated by the Carnegie Institution for Science. The SDSS website is \url{www.sdss.org}.

SDSS is managed by the Astrophysical Research Consortium for the Participating Institutions of the SDSS Collaboration, including Caltech, The Carnegie Institution for Science, Chilean National Time Allocation Committee (CNTAC) ratified researchers, The Flatiron Institute, the Gotham Participation Group, Harvard University, Heidelberg University, The Johns Hopkins University, L’Ecole polytechnique f\'{e}d\'{e}rale de Lausanne (EPFL), Leibniz-Institut f\"{u}r Astrophysik Potsdam (AIP), Max-Planck-Institut f\"{u}r Astronomie (MPIA Heidelberg), Max-Planck-Institut f\"{u}r Extraterrestrische Physik (MPE), Nanjing University, National Astronomical Observatories of China (NAOC), New Mexico State University, The Ohio State University, Pennsylvania State University, Smithsonian Astrophysical Observatory, Space Telescope Science Institute (STScI), the Stellar Astrophysics Participation Group, Universidad Nacional Aut\'{o}noma de M\'{e}xico, University of Arizona, University of Colorado Boulder, University of Illinois at Urbana-Champaign, University of Toronto, University of Utah, University of Virginia, Yale University, and Yunnan University.

This work has made use of data from the European Space Agency (ESA) mission
{\it Gaia} (\url{https://www.cosmos.esa.int/gaia}), processed by the {\it Gaia}
Data Processing and Analysis Consortium (DPAC,
\url{https://www.cosmos.esa.int/web/gaia/dpac/consortium}). Funding for the DPAC
has been provided by national institutions, in particular the institutions
participating in the {\it Gaia} Multilateral Agreement.
\end{acknowledgments}

\bibliography{snc_sf}{}

@ARTICLE{rix2021,
       author = {{Rix}, Hans-Walter and {Hogg}, David W. and {Boubert}, Douglas and {Brown}, Anthony G.~A. and {Casey}, Andrew and {Drimmel}, Ronald and {Everall}, Andrew and {Fouesneau}, Morgan and {Price-Whelan}, Adrian M.},
        title = "{Selection Functions in Astronomical Data Modeling, with the Space Density of White Dwarfs as a Worked Example}",
      journal = {\aj},
         year = 2021,
        month = oct,
       volume = {162},
       number = {4},
          eid = {142},
        pages = {142},
          doi = {10.3847/1538-3881/ac0c13},
archivePrefix = {arXiv},
       eprint = {2106.07653},
 primaryClass = {astro-ph.IM},
       adsurl = {https://ui.adsabs.harvard.edu/abs/2021AJ....162..142R}
}

@ARTICLE{subsamplegaia,
       author = {{Castro-Ginard}, Alfred and {Brown}, Anthony G.~A. and {Kostrzewa-Rutkowska}, Zuzanna and {Cantat-Gaudin}, Tristan and {Drimmel}, Ronald and {Oh}, Semyeong and {Belokurov}, Vasily and {Casey}, Andrew R. and {Fouesneau}, Morgan and {Khanna}, Shourya and {Price-Whelan}, Adrian M. and {Rix}, Hans-Walter},
        title = "{Estimating the selection function of Gaia DR3 subsamples}",
      journal = {\aap},
         year = 2023,
        month = sep,
       volume = {677},
          eid = {A37},
        pages = {A37},
          doi = {10.1051/0004-6361/202346547},
archivePrefix = {arXiv},
       eprint = {2303.17738},
 primaryClass = {astro-ph.GA},
       adsurl = {https://ui.adsabs.harvard.edu/abs/2023A&A...677A..37C}
}

@ARTICLE{gaiasf,
       author = {{Cantat-Gaudin}, Tristan and {Fouesneau}, Morgan and {Rix}, Hans-Walter and {Brown}, Anthony G.~A. and {Castro-Ginard}, Alfred and {Kostrzewa-Rutkowska}, Zuzanna and {Drimmel}, Ronald and {Hogg}, David W. and {Casey}, Andrew R. and {Khanna}, Shourya and {Oh}, Semyeong and {Price-Whelan}, Adrian M. and {Belokurov}, Vasily and {Saydjari}, Andrew K. and {Green}, G.},
        title = "{An empirical model of the Gaia DR3 selection function}",
      journal = {\aap},
         year = 2023,
        month = jan,
       volume = {669},
          eid = {A55},
        pages = {A55},
          doi = {10.1051/0004-6361/202244784},
archivePrefix = {arXiv},
       eprint = {2208.09335},
 primaryClass = {astro-ph.GA},
       adsurl = {https://ui.adsabs.harvard.edu/abs/2023A&A...669A..55C}
}

@ARTICLE{Schmidt1968,
       author = {{Schmidt}, Maarten},
        title = "{Space Distribution and Luminosity Functions of Quasi-Stellar Radio Sources}",
      journal = {\apj},
         year = 1968,
        month = feb,
       volume = {151},
        pages = {393},
          doi = {10.1086/149446},
       adsurl = {https://ui.adsabs.harvard.edu/abs/1968ApJ...151..393S}
}

@ARTICLE{felten1976,
       author = {{Felten}, J.~E.},
        title = "{On Schmidt's V ?m estimator and other estimators of luminosity functions.}",
      journal = {\apj},
         year = 1976,
        month = aug,
       volume = {207},
        pages = {700-709},
          doi = {10.1086/154538},
       adsurl = {https://ui.adsabs.harvard.edu/abs/1976ApJ...207..700F}
}

@ARTICLE{tinney1993,
       author = {{Tinney}, C.~G. and {Reid}, I.~N. and {Mould}, J.~R.},
        title = "{The Faintest Stars: From Schmidt Plates to Luminosity Functions}",
      journal = {\apj},
         year = 1993,
        month = sep,
       volume = {414},
        pages = {254},
          doi = {10.1086/173074},
       adsurl = {https://ui.adsabs.harvard.edu/abs/1993ApJ...414..254T}
}

@ARTICLE{GNNS,
       author = {{Gaia Collaboration} and {Smart}, R.~L. and {Sarro}, L.~M. and {Rybizki}, J. and {Reyl{\'e}}, C. and {Robin}, A.~C. and {Hambly}, N.~C. and {Abbas}, U. and {Barstow}, M.~A. and {de Bruijne}, J.~H.~J. and {Bucciarelli}, B. and {Carrasco}, J.~M. and {Cooper}, W.~J. and {Hodgkin}, S.~T. and {Masana}, E. and {Michalik}, D. and {Sahlmann}, J. and {Sozzetti}, A. and {Brown}, A.~G.~A. and {Vallenari}, A. and {Prusti}, T. and {Babusiaux}, C. and {Biermann}, M. and {Creevey}, O.~L. and {Evans}, D.~W. and {Eyer}, L. and {Hutton}, A. and {Jansen}, F. and {Jordi}, C. and {Klioner}, S.~A. and {Lammers}, U. and {Lindegren}, L. and {Luri}, X. and {Mignard}, F. and {Panem}, C. and {Pourbaix}, D. and {Randich}, S. and {Sartoretti}, P. and {Soubiran}, C. and {Walton}, N.~A. and {Arenou}, F. and {Bailer-Jones}, C.~A.~L. and {Bastian}, U. and {Cropper}, M. and {Drimmel}, R. and {Katz}, D. and {Lattanzi}, M.~G. and {van Leeuwen}, F. and {Bakker}, J. and {Casta{\~n}eda}, J. and {De Angeli}, F. and {Ducourant}, C. and {Fabricius}, C. and {Fouesneau}, M. and {Fr{\'e}mat}, Y. and {Guerra}, R. and {Guerrier}, A. and {Guiraud}, J. and {Jean-Antoine Piccolo}, A. and {Messineo}, R. and {Mowlavi}, N. and {Nicolas}, C. and {Nienartowicz}, K. and {Pailler}, F. and {Panuzzo}, P. and {Riclet}, F. and {Roux}, W. and {Seabroke}, G.~M. and {Sordo}, R. and {Tanga}, P. and {Th{\'e}venin}, F. and {Gracia-Abril}, G. and {Portell}, J. and {Teyssier}, D. and {Altmann}, M. and {Andrae}, R. and {Bellas-Velidis}, I. and {Benson}, K. and {Berthier}, J. and {Blomme}, R. and {Brugaletta}, E. and {Burgess}, P.~W. and {Busso}, G. and {Carry}, B. and {Cellino}, A. and {Cheek}, N. and {Clementini}, G. and {Damerdji}, Y. and {Davidson}, M. and {Delchambre}, L. and {Dell'Oro}, A. and {Fern{\'a}ndez-Hern{\'a}ndez}, J. and {Galluccio}, L. and {Garc{\'\i}a-Lario}, P. and {Garcia-Reinaldos}, M. and {Gonz{\'a}lez-N{\'u}{\~n}ez}, J. and {Gosset}, E. and {Haigron}, R. and {Halbwachs}, J. -L. and {Harrison}, D.~L. and {Hatzidimitriou}, D. and {Heiter}, U. and {Hern{\'a}ndez}, J. and {Hestroffer}, D. and {Holl}, B. and {Jan{\ss}en}, K. and {Jevardat de Fombelle}, G. and {Jordan}, S. and {Krone-Martins}, A. and {Lanzafame}, A.~C. and {L{\"o}ffler}, W. and {Lorca}, A. and {Manteiga}, M. and {Marchal}, O. and {Marrese}, P.~M. and {Moitinho}, A. and {Mora}, A. and {Muinonen}, K. and {Osborne}, P. and {Pancino}, E. and {Pauwels}, T. and {Recio-Blanco}, A. and {Richards}, P.~J. and {Riello}, M. and {Rimoldini}, L. and {Roegiers}, T. and {Siopis}, C. and {Smith}, M. and {Ulla}, A. and {Utrilla}, E. and {van Leeuwen}, M. and {van Reeven}, W. and {Abreu Aramburu}, A. and {Accart}, S. and {Aerts}, C. and {Aguado}, J.~J. and {Ajaj}, M. and {Altavilla}, G. and {{\'A}lvarez}, M.~A. and {{\'A}lvarez Cid-Fuentes}, J. and {Alves}, J. and {Anderson}, R.~I. and {Anglada Varela}, E. and {Antoja}, T. and {Audard}, M. and {Baines}, D. and {Baker}, S.~G. and {Balaguer-N{\'u}{\~n}ez}, L. and {Balbinot}, E. and {Balog}, Z. and {Barache}, C. and {Barbato}, D. and {Barros}, M. and {Bartolom{\'e}}, S. and {Bassilana}, J. -L. and {Bauchet}, N. and {Baudesson-Stella}, A. and {Becciani}, U. and {Bellazzini}, M. and {Bernet}, M. and {Bertone}, S. and {Bianchi}, L. and {Blanco-Cuaresma}, S. and {Boch}, T. and {Bombrun}, A. and {Bossini}, D. and {Bouquillon}, S. and {Bragaglia}, A. and {Bramante}, L. and {Breedt}, E. and {Bressan}, A. and {Brouillet}, N. and {Burlacu}, A. and {Busonero}, D. and {Butkevich}, A.~G. and {Buzzi}, R. and {Caffau}, E. and {Cancelliere}, R. and {C{\'a}novas}, H. and {Cantat-Gaudin}, T. and {Carballo}, R. and {Carlucci}, T. and {Carnerero}, M.~I. and {Casamiquela}, L. and {Castellani}, M. and {Castro-Ginard}, A. and {Castro Sampol}, P. and {Chaoul}, L. and {Charlot}, P. and {Chemin}, L. and {Chiavassa}, A. and {Cioni}, M. -R.~L. and {Comoretto}, G. and {Cornez}, T. and {Cowell}, S. and {Crifo}, F. and {Crosta}, M. and {Crowley}, C. and {Dafonte}, C. and {Dapergolas}, A.},
        title = "{Gaia Early Data Release 3. The Gaia Catalogue of Nearby Stars}",
      journal = {\aap},
         year = 2021,
        month = may,
       volume = {649},
          eid = {A6},
        pages = {A6},
          doi = {10.1051/0004-6361/202039498},
archivePrefix = {arXiv},
       eprint = {2012.02061},
 primaryClass = {astro-ph.SR},
       adsurl = {https://ui.adsabs.harvard.edu/abs/2021A&A...649A...6G}
}

@ARTICLE{RECONS,
       author = {{Henry}, Todd J. and {Jao}, Wei-Chun and {Winters}, Jennifer G. and {Dieterich}, Sergio B. and {Finch}, Charlie T. and {Ianna}, Philip A. and {Riedel}, Adric R. and {Silverstein}, Michele L. and {Subasavage}, John P. and {Vrijmoet}, Eliot Halley},
        title = "{The Solar Neighborhood XLIV: RECONS Discoveries within 10 parsecs}",
      journal = {\aj},
         year = 2018,
        month = jun,
       volume = {155},
       number = {6},
          eid = {265},
        pages = {265},
          doi = {10.3847/1538-3881/aac262},
archivePrefix = {arXiv},
       eprint = {1804.07377},
 primaryClass = {astro-ph.SR},
       adsurl = {https://ui.adsabs.harvard.edu/abs/2018AJ....155..265H}
}

@ARTICLE{sdssV,
        author = {{Kollmeier}, Juna A. and {Rix}, Hans-Walter and {Aerts}, Conny and {Aird}, James and {Vera Alfaro}, Pablo and {Almeida}, Andr{\'e}s and {Anderson}, Scott F. and {Arseneau}, Stefan M. and {Assef}, Roberto J. and {Aviram}, Shir and {Aydar}, Catarina and {Badenes}, Carles and {Bandyopadhyay}, Avrajit and {Barger}, Kat and {Barkhouser}, Robert H. and {Bauer}, Franz E. and {Behmard}, Aida and {Bender}, Chad and {Besser}, Felipe and {Bhattarai}, Binod and {Bilgi}, Pavaman and {Bird}, Jonathan and {Bizyaev}, Dmitry and {Blanc}, Guillermo A. and {Blanton}, Michael R. and {Bochanski}, John and {Bovy}, Jo and {Brandon}, Christopher and {Brandt}, William Nielsen and {Brownstein}, Joel R. and {Buchner}, Johannes and {Burchett}, Joseph N. and {Carlberg}, Joleen and {Casey}, Andrew R. and {Castaneda-Carlos}, Lesly and {Chakraborty}, Priyanka and {Chanam{\'e}}, Julio and {Chandra}, Vedant and {Cherinka}, Brian and {Chilingarian}, Igor and {Comparat}, Johan and {Cosens}, Maren and {Covey}, Kevin and {Crane}, Jeffrey D. and {Crumpler}, Nicole R. and {Cruz-Gonzalez}, Irene and {Cunha}, Katia and {Cunningham}, Tim and {Dai}, Xinyu and {Darling}, Jeremy and {Davidson}, Jr., James W. and {Davis}, Megan C. and {De Lee}, Nathan and {Deacon}, Niall and {M{\'e}ndez Delgado}, Jos{\'e} Eduardo and {Demasi}, Sebastian and {Demianenko}, Mariia and {Derwent}, Mark and {D'Onghia}, Elena and {Di Mille}, Francesco and {Dias}, Bruno and {Donor}, John and {Dow}, Peter N. and {Drory}, Niv and {Dwelly}, Tom and {Egorov}, Oleg and {Egorova}, Evgeniya and {El-Badry}, Kareem and {Engelman}, Mike and {Eracleous}, Mike and {Fan}, Xiaohui and {Farr}, Emily and {Fries}, Logan and {Frinchaboy}, Peter and {Froning}, Cynthia S. and {G{\"a}nsicke}, Boris T. and {Garc{\'\i}a}, Pablo and {Gelfand}, Joseph and {Gentile Fusillo}, Nicola Pietro and {Glover}, Simon and {Grabowski}, Katie and {Grebel}, Eva K. and {Green}, Paul J. and {Grier}, Catherine and {Gupta}, Pramod and {Gray}, Aidan C. and {H{\"a}berle}, Maximilian and {Hall}, Patrick B. and {Hammond}, Randolph P. and {Hawkins}, Keith and {Harding}, Albert C. and {Heged{\H{u}}s}, Viola and {Herbst}, Tom and {Hermes}, J.~J. and {Rodr{\'\i}guez Hidalgo}, Paola and {Hilder}, Thomas and {Hogg}, David W. and {Holtzman}, Jon A. and {Horta}, Danny and {Huang}, Yang and {Hwang}, Hsiang-Chih and {Ibarra-Medel}, Hector Javier and {Imig}, Julie and {Inight}, Keith and {Jana}, Arghajit and {Ji}, Alexander P. and {Jim{\'e}nez-Arranz}, {\'O}scar and {Jofre}, Paula and {Johns}, Matt and {Johnson}, Jennifer and {Johnson}, James W. and {Johnston}, Evelyn J. and {Jones}, Amy M. and {Katkov}, Ivan and {Knapp}, Gillian R. and {Koekemoer}, Anton M. and {Kounkel}, Marina and {Kreckel}, Kathryn and {Krishnarao}, Dhanesh and {Krumpe}, Mirko and {Kumari}, Nimisha and {Kupfer}, Thomas and {Lacerna}, Ivan and {Laporte}, Chervin and {Lepine}, Sebastien and {Li}, Jing and {Liu}, Xin and {Loebman}, Sarah and {Long}, Knox and {Roman-Lopes}, Alexandre and {Lu}, Yuxi and {Majewski}, Steven Raymond and {Maoz}, Dan and {McKinnon}, Kevin A. and {Medan}, Ilija and {Merloni}, Andrea and {Minniti}, Dante and {Morrison}, Sean and {Myers}, Natalie and {M{\'e}sz{\'a}ros}, Szabolcs and {Nandra}, Kirpal and {Nayak}, Prasanta K. and {Ness}, Melissa K. and {Nidever}, David L. and {O'Brien}, Thomas and {Oeur}, Micah and {Oravetz}, Audrey and {Oravetz}, Daniel and {Otto}, Jonah and {Pallathadka}, Gautham Adamane and {Palunas}, Povilas and {Pan}, Kaike and {Pappalardo}, Daniel and {Pandey}, Rakesh and {Negrete Pe{\~n}aloza}, Castalia Alenka and {Pinsonneault}, Marc H. and {Pogge}, Richard W. and {Taghizadeh Popp}, Manuchehr and {Price-Whelan}, Adrian M. and {Pulatova}, Nadiia and {Qiu}, Dan and {Ramirez}, Solange and {Rankine}, Amy and {Ricci}, Claudio and {Runnoe}, Jessie C. and {Sanchez}, Sebastian and {Salvato}, Mara and {Sarbadhicary}, Sumit K. and {Sattler}, Natascha and {Saydjari}, Andrew K. and {Sayres}, Conor and {Schinnerer}, Eva and {Schlaufman}, Kevin C. and {Schneider}, Donald P. and {Schreiber}, Matthias R. and {Schwope}, Axel and {Serna}, Javier and {Shen}, Yue and {Sif{\'o}n}, Crist{\'o}bal and {Singh}, Amrita and {Sinha}, Amaya and {Smee}, Stephen and {Song}, Ying-Yi and {Souto}, Diogo and {Stassun}, Keivan G. and {Steinmetz}, Matthias and {Stone-Martinez}, Alexander and {Stringfellow}, Guy and {Stutz}, Amelia and {S{\'a}nchez-Gallego}, Jos{\'e} and {Tan}, Jonathan C. and {Tayar}, Jamie and {Thai}, Riley and {Thakar}, Ani and {Ting}, Yuan-Sen and {Tkachenko}, Andrew and {Tovmassian}, Gagik and {Trakhtenbrot}, Benny and {Fern{\'a}ndez-Trincado}, Jos{\'e} G. and {Troup}, Nicholas},
        title = "{Sloan Digital Sky Survey. V. Pioneering Panoptic Spectroscopy}",
      journal = {\aj},
         year = 2026,
        month = jan,
       volume = {171},
       number = {1},
          eid = {52},
        pages = {52},
          doi = {10.3847/1538-3881/ae0576},
archivePrefix = {arXiv},
       eprint = {2507.06989},
 primaryClass = {astro-ph.IM},
       adsurl = {https://ui.adsabs.harvard.edu/abs/2026AJ....171...52K}
}

@ARTICLE{APO,
       author = {{Gunn}, James E. and {Siegmund}, Walter A. and {Mannery}, Edward J. and {Owen}, Russell E. and {Hull}, Charles L. and {Leger}, R. French and {Carey}, Larry N. and {Knapp}, Gillian R. and {York}, Donald G. and {Boroski}, William N. and {Kent}, Stephen M. and {Lupton}, Robert H. and {Rockosi}, Constance M. and {Evans}, Michael L. and {Waddell}, Patrick and {Anderson}, John E. and {Annis}, James and {Barentine}, John C. and {Bartoszek}, Larry M. and {Bastian}, Steven and {Bracker}, Stephen B. and {Brewington}, Howard J. and {Briegel}, Charles I. and {Brinkmann}, Jon and {Brown}, Yorke J. and {Carr}, Michael A. and {Czarapata}, Paul C. and {Drennan}, Craig C. and {Dombeck}, Thomas and {Federwitz}, Glenn R. and {Gillespie}, Bruce A. and {Gonzales}, Carlos and {Hansen}, Sten U. and {Harvanek}, Michael and {Hayes}, Jeffrey and {Jordan}, Wendell and {Kinney}, Ellyne and {Klaene}, Mark and {Kleinman}, S.~J. and {Kron}, Richard G. and {Kresinski}, Jurek and {Lee}, Glenn and {Limmongkol}, Siriluk and {Lindenmeyer}, Carl W. and {Long}, Daniel C. and {Loomis}, Craig L. and {McGehee}, Peregrine M. and {Mantsch}, Paul M. and {Neilsen}, Eric H., Jr. and {Neswold}, Richard M. and {Newman}, Peter R. and {Nitta}, Atsuko and {Peoples}, John, Jr. and {Pier}, Jeffrey R. and {Prieto}, Peter S. and {Prosapio}, Angela and {Rivetta}, Claudio and {Schneider}, Donald P. and {Snedden}, Stephanie and {Wang}, Shu-i.},
        title = "{The 2.5 m Telescope of the Sloan Digital Sky Survey}",
      journal = {\aj},
         year = 2006,
        month = apr,
       volume = {131},
       number = {4},
        pages = {2332-2359},
          doi = {10.1086/500975},
archivePrefix = {arXiv},
       eprint = {astro-ph/0602326},
 primaryClass = {astro-ph},
       adsurl = {https://ui.adsabs.harvard.edu/abs/2006AJ....131.2332G}
}

@ARTICLE{LCO,
       author = {{Bowen}, I.~S. and {Vaughan}, A.~H., Jr.},
        title = "{The optical design of the 40-in. telescope and of the Ir{\'e}n{\'e}e DuPont telescope at Las Campanas Observatory, Chile.}",
      journal = {\ao},
         year = 1973,
        month = jan,
       volume = {12},
        pages = {1430-1434},
          doi = {10.1364/AO.12.001430},
       adsurl = {https://ui.adsabs.harvard.edu/abs/1973ApOpt..12.1430B}
}

@INPROCEEDINGS{FPS,
       author = {{Pogge}, Richard W. and {Derwent}, Mark A. and {O'Brien}, Thomas P. and {Jurgenson}, Colby A. and {Pappalardo}, Daniel and {Engelman}, Michael and {Brandon}, Christopher and {Brady}, Julia and {Clawson}, Nicholas and {Shover}, Jon and {Mason}, Jerry and {Kneib}, Jean-Paul and {Araujo}, Ricardo and {Bouri}, Mohamed and {Kronig}, Luzius and {Grossen}, Lo{\"\i}c. and {Gillet}, Denis and {Macktoobian}, Matin and {Tuttle}, Sarah E. and {Farr}, Emily and {S{\'a}nchez-Gallego}, Jos{\'e} and {Sayres}, Conor},
        title = "{A robotic Focal Plane System (FPS) for the Sloan Digital Sky Survey V}",
    booktitle = {Ground-based and Airborne Instrumentation for Astronomy VIII},
         year = 2020,
       editor = {{Evans}, Christopher J. and {Bryant}, Julia J. and {Motohara}, Kentaro},
       series = {Society of Photo-Optical Instrumentation Engineers (SPIE) Conference Series},
       volume = {11447},
        month = dec,
          eid = {1144781},
        pages = {1144781},
          doi = {10.1117/12.2561113},
       adsurl = {https://ui.adsabs.harvard.edu/abs/2020SPIE11447E..81P}
}

@ARTICLE{blanton2025,
       author = {{Blanton}, Michael R. and {Carlberg}, Joleen K. and {Dwelly}, Tom and {Medan}, Ilija and {Chojnowski}, S. Drew and {Covey}, Kevin and {Davis}, Megan C. and {Donor}, John and {Gupta}, Pramod and {Ji}, Alexander and {Johnson}, Jennifer A. and {Kollmeier}, Juna A. and {Sanchez-Gallego}, Jose and {Sayres}, Conor and {Zari}, Eleonora},
        title = "{robostrategy: Field and Target Assignment Optimization in the Sloan Digital Sky Survey V}",
      journal = {arXiv e-prints},
         year = 2025,
        month = may,
          eid = {arXiv:2505.21328},
        pages = {arXiv:2505.21328},
          doi = {10.48550/arXiv.2505.21328},
archivePrefix = {arXiv},
       eprint = {2505.21328},
 primaryClass = {astro-ph.IM},
       adsurl = {https://ui.adsabs.harvard.edu/abs/2025arXiv250521328B}
}

@ARTICLE{medan2025,
       author = {{Medan}, Ilija and {Dwelly}, Tom and {Covey}, Kevin R. and {Zari}, Eleonora and {Blanton}, Michael R. and {Carlberg}, Joleen K. and {Chojnowski}, S. Drew and {Ji}, Alexander and {Shen}, Yue and {Donor}, John and {S{\'a}nchez-Gallego}, Jos{\'e} and {Morrison}, Sean and {Ibarra-Medel}, H{\'e}ctor J. and {Sayres}, Conor and {Stassun}, Keivan G.},
        title = "{Procedures for Constraining Robotic Fiber Positioning for Highly Multiplexed Spectroscopic Surveys: The Case of FPS for SDSS-V}",
      journal = {arXiv e-prints},
         year = 2025,
        month = jun,
          eid = {arXiv:2506.15475},
        pages = {arXiv:2506.15475},
          doi = {10.48550/arXiv.2506.15475},
archivePrefix = {arXiv},
       eprint = {2506.15475},
 primaryClass = {astro-ph.IM},
       adsurl = {https://ui.adsabs.harvard.edu/abs/2025arXiv250615475M}
}

@ARTICLE{sdssdr18,
       author = {{Almeida}, Andr{\'e}s and {Anderson}, Scott F. and {Argudo-Fern{\'a}ndez}, Maria and {Badenes}, Carles and {Barger}, Kat and {Barrera-Ballesteros}, Jorge K. and {Bender}, Chad F. and {Benitez}, Erika and {Besser}, Felipe and {Bird}, Jonathan C. and {Bizyaev}, Dmitry and {Blanton}, Michael R. and {Bochanski}, John and {Bovy}, Jo and {Brandt}, William Nielsen and {Brownstein}, Joel R. and {Buchner}, Johannes and {Bulbul}, Esra and {Burchett}, Joseph N. and {Cano D{\'\i}az}, Mariana and {Carlberg}, Joleen K. and {Casey}, Andrew R. and {Chandra}, Vedant and {Cherinka}, Brian and {Chiappini}, Cristina and {Coker}, Abigail A. and {Comparat}, Johan and {Conroy}, Charlie and {Contardo}, Gabriella and {Cortes}, Arlin and {Covey}, Kevin and {Crane}, Jeffrey D. and {Cunha}, Katia and {Dabbieri}, Collin and {Davidson}, James W. and {Davis}, Megan C. and {de Andrade Queiroz}, Anna Barbara and {De Lee}, Nathan and {M{\'e}ndez Delgado}, Jos{\'e} Eduardo and {Demasi}, Sebastian and {Di Mille}, Francesco and {Donor}, John and {Dow}, Peter and {Dwelly}, Tom and {Eracleous}, Mike and {Eriksen}, Jamey and {Fan}, Xiaohui and {Farr}, Emily and {Frederick}, Sara and {Fries}, Logan and {Frinchaboy}, Peter and {G{\"a}nsicke}, Boris T. and {Ge}, Junqiang and {Gonz{\'a}lez {\'A}vila}, Consuelo and {Grabowski}, Katie and {Grier}, Catherine and {Guiglion}, Guillaume and {Gupta}, Pramod and {Hall}, Patrick and {Hawkins}, Keith and {Hayes}, Christian R. and {Hermes}, J.~J. and {Hern{\'a}ndez-Garc{\'\i}a}, Lorena and {Hogg}, David W. and {Holtzman}, Jon A. and {Ibarra-Medel}, Hector Javier and {Ji}, Alexander and {Jofre}, Paula and {Johnson}, Jennifer A. and {Jones}, Amy M. and {Kinemuchi}, Karen and {Kluge}, Matthias and {Koekemoer}, Anton and {Kollmeier}, Juna A. and {Kounkel}, Marina and {Krishnarao}, Dhanesh and {Krumpe}, Mirko and {Lacerna}, Ivan and {Lago}, Paulo Jakson Assuncao and {Laporte}, Chervin and {Liu}, Chao and {Liu}, Ang and {Liu}, Xin and {Lopes}, Alexandre Roman and {Macktoobian}, Matin and {Majewski}, Steven R. and {Malanushenko}, Viktor and {Maoz}, Dan and {Masseron}, Thomas and {Masters}, Karen L. and {Matijevic}, Gal and {McBride}, Aidan and {Medan}, Ilija and {Merloni}, Andrea and {Morrison}, Sean and {Myers}, Natalie and {M{\'e}sz{\'a}ros}, Szabolcs and {Negrete}, C. Alenka and {Nidever}, David L. and {Nitschelm}, Christian and {Oravetz}, Daniel and {Oravetz}, Audrey and {Pan}, Kaike and {Peng}, Yingjie and {Pinsonneault}, Marc H. and {Pogge}, Rick and {Qiu}, Dan and {Ramirez}, Solange V. and {Rix}, Hans-Walter and {Fern{\'a}ndez Rosso}, Daniela and {Runnoe}, Jessie and {Salvato}, Mara and {Sanchez}, Sebastian F. and {Santana}, Felipe A. and {Saydjari}, Andrew and {Sayres}, Conor and {Schlaufman}, Kevin C. and {Schneider}, Donald P. and {Schwope}, Axel and {Serna}, Javier and {Shen}, Yue and {Sobeck}, Jennifer and {Song}, Ying-Yi and {Souto}, Diogo and {Spoo}, Taylor and {Stassun}, Keivan G. and {Steinmetz}, Matthias and {Straumit}, Ilya and {Stringfellow}, Guy and {S{\'a}nchez-Gallego}, Jos{\'e} and {Taghizadeh-Popp}, Manuchehr and {Tayar}, Jamie and {Thakar}, Ani and {Tissera}, Patricia B. and {Tkachenko}, Andrew and {Hernandez Toledo}, Hector and {Trakhtenbrot}, Benny and {Fern{\'a}ndez-Trincado}, Jos{\'e} G. and {Troup}, Nicholas and {Trump}, Jonathan R. and {Tuttle}, Sarah and {Ulloa}, Natalie and {Vazquez-Mata}, Jose Antonio and {Vera Alfaro}, Pablo and {Villanova}, Sandro and {Wachter}, Stefanie and {Weijmans}, Anne-Marie and {Wheeler}, Adam and {Wilson}, John and {Wojno}, Leigh and {Wolf}, Julien and {Xue}, Xiang-Xiang and {Ybarra}, Jason E. and {Zari}, Eleonora and {Zasowski}, Gail},
        title = "{The Eighteenth Data Release of the Sloan Digital Sky Surveys: Targeting and First Spectra from SDSS-V}",
      journal = {\apjs},
         year = 2023,
        month = aug,
       volume = {267},
       number = {2},
          eid = {44},
        pages = {44},
          doi = {10.3847/1538-4365/acda98},
archivePrefix = {arXiv},
       eprint = {2301.07688},
 primaryClass = {astro-ph.GA},
       adsurl = {https://ui.adsabs.harvard.edu/abs/2023ApJS..267...44A}
}

@article{ smee13a,
author = {Smee, Stephen A and Gunn, James E and Uomoto, Alan and Roe, 
Natalie and Schlegel, David and Rockosi, Constance M and Carr, Michael A
 and Leger, French and Dawson, Kyle S and Olmstead, Matthew D and 
Brinkmann, Jon and Owen, Russell and Barkhouser, Robert H and Honscheid,
 Klaus and Harding, Paul and Long, Dan and Lupton, Robert H and Loomis, 
Craig and Anderson, Lauren and Annis, James and Bernardi, Marangela and 
Bhardwaj, Vaishali and Bizyaev, Dmitry and bolton, Adam S and 
Brewington, Howard and Briggs, John W and Burles, Scott and Burns, James
 G and Castander, Francisco Javier and Connolly, Andrew and Davenport, 
James R A and Ebelke, Garrett and Epps, Harland and Feldman, Paul D and 
Friedman, Scott D and Frieman, Joshua and Heckman, Timothy and Hull, 
Charles L and Knapp, Gillian R and Lawrence, David M and Loveday, Jon 
and Mannery, Edward J and Malanushenko, Elena and Malanushenko, Viktor 
and Merrelli, Aronne James and Muna, Demitri and Newman, Peter R and 
Nichol, Robert C and Oravetz, Daniel and Pan, Kaike and Pope, Adrian C 
and Ricketts, Paul G and Shelden, Alaina and Sandford, Dale and 
Siegmund, Walter and Simmons, Audrey and Smith, D Shane and Snedden, 
Stephanie and Schneider, Donald P and SubbaRao, Mark and Tremonti, 
Christy and Waddell, Patrick and York, Donald G},
title = {{The Multi-object, Fiber-fed Spectrographs for the Sloan 
Digital Sky Survey and the Baryon Oscillation Spectroscopic Survey}},
journal = {The Astronomical Journal},
year = {2013},
volume = {146},
number = {2},
pages = {32},
month = aug
}

@ARTICLE{ wilson19a,
       author = {{Wilson}, J.~C. and {Hearty}, F.~R. and {Skrutskie}, M.~F. and
         {Majewski}, S.~R. and {Holtzman}, J.~A. and {Eisenstein}, D. and
         {Gunn}, J. and {Blank}, B. and {Henderson}, C. and {Smee}, S. and
         {Nelson}, M. and {Nidever}, D. and {Arns}, J. and {Barkhouser}, R. and
         {Barr}, J. and {Beland}, S. and {Bershady}, M.~A. and {Blanton}, M.~R. and
         {Brunner}, S. and {Burton}, A. and {Carey}, L. and {Carr}, M. and
         {Colque}, J.~P. and {Crane}, J. and {Damke}, G.~J. and
         {Davidson}, J.~W., Jr. and {Dean}, J. and {Di Mille}, F. and
         {Don}, K.~W. and {Ebelke}, G. and {Evans}, M. and {Fitzgerald}, G. and
         {Gillespie}, B. and {Hall}, M. and {Harding}, A. and {Harding}, P. and
         {Hammond}, R. and {Hancock}, D. and {Harrison}, C. and {Hope}, S. and
         {Horne}, T. and {Karakla}, J. and {Lam}, C. and {Leger}, F. and
         {MacDonald}, N. and {Maseman}, P. and {Matsunari}, J. and {Melton}, S. and
         {Mitcheltree}, T. and {O'Brien}, T. and {O'Connell}, R.~W. and
         {Patten}, A. and {Richardson}, W. and {Rieke}, G. and {Rieke}, M. and
         {Roman-Lopes}, A. and {Schiavon}, R.~P. and {Sobeck}, J.~S. and
         {Stolberg}, T. and {Stoll}, R. and {Tembe}, M. and {Trujillo}, J.~D. and
         {Uomoto}, A. and {Vernieri}, M. and {Walker}, E. and {Weinberg}, D.~H. and
         {Young}, E. and {Anthony-Brumfield}, B. and {Bizyaev}, D. and
         {Breslauer}, B. and {De Lee}, N. and {Downey}, J. and {Halverson}, S. and
         {Huehnerhoff}, J. and {Klaene}, M. and {Leon}, E. and {Long}, D. and
         {Mahadevan}, S. and {Malanushenko}, E. and {Nguyen}, D.~C. and
         {Owen}, R. and {S{\'a}nchez-Gallego}, J.~R. and {Sayres}, C. and
         {Shane}, N. and {Shectman}, S.~A. and {Shetrone}, M. and {Skinner}, D. and
         {Stauffer}, F. and {Zhao}, B.},
        title = "{The Apache Point Observatory Galactic Evolution Experiment (APOGEE) Spectrographs}",
      journal = {\pasp},
         year = 2019,
        month = may,
       volume = {131},
       number = {999},
        pages = {055001},
          doi = {10.1088/1538-3873/ab0075},
archivePrefix = {arXiv},
       eprint = {1902.00928},
 primaryClass = {astro-ph.IM},
       adsurl = {https://ui.adsabs.harvard.edu/abs/2019PASP..131e5001W}
}

@software{jax2018github,
  author = {James Bradbury and Roy Frostig and Peter Hawkins and Matthew James Johnson and Chris Leary and Dougal Maclaurin and George Necula and Adam Paszke and Jake Vander{P}las and Skye Wanderman-{M}ilne and Qiao Zhang},
  title = {{JAX}: composable transformations of {P}ython+{N}um{P}y programs},
  url = {http://github.com/jax-ml/jax},
  version = {0.3.13},
  year = {2018},
}

@article{numpyro,
  title={Composable Effects for Flexible and Accelerated Probabilistic Programming in NumPyro},
  author={Phan, Du and Pradhan, Neeraj and Jankowiak, Martin},
  journal={arXiv preprint arXiv:1912.11554},
  year={2019}
}

@ARTICLE{lutz1973,
       author = {{Lutz}, Thomas E. and {Kelker}, Douglas H.},
        title = "{On the Use of Trigonometric Parallaxes for the Calibration of Luminosity Systems: Theory}",
      journal = {\pasp},
         year = 1973,
        month = oct,
       volume = {85},
       number = {507},
        pages = {573},
          doi = {10.1086/129506},
       adsurl = {https://ui.adsabs.harvard.edu/abs/1973PASP...85..573L}
}

@ARTICLE{lineforest,
       author = {{Saad}, Serat and {Lane}, Kaitlyn and {Kounkel}, Marina and {Stassun}, Keivan G. and {L{\'o}pez-Valdivia}, Ricardo and {Kim}, Jinyoung Serena and {Pe{\~n}a Ram{\'\i}rez}, Karla and {Stringfellow}, Guy S. and {Rom{\'a}n-Z{\'u}{\~n}iga}, Carlos G. and {Hern{\'a}ndez}, Jes{\'u}s and {Wolk}, Scott J. and {Hillenbrand}, Lynne A.},
        title = "{ABYSS. II. Identification of Young Stars in Optical SDSS Spectra and Their Properties}",
      journal = {\aj},
         year = 2024,
        month = mar,
       volume = {167},
       number = {3},
          eid = {125},
        pages = {125},
          doi = {10.3847/1538-3881/ad2001},
archivePrefix = {arXiv},
       eprint = {2401.01932},
 primaryClass = {astro-ph.SR},
       adsurl = {https://ui.adsabs.harvard.edu/abs/2024AJ....167..125S}
}

@ARTICLE{NUTS,
       author = {{Hoffman}, Matthew D. and {Gelman}, Andrew},
        title = "{The No-U-Turn Sampler: Adaptively Setting Path Lengths in Hamiltonian Monte Carlo}",
      journal = {arXiv e-prints},
         year = 2011,
        month = nov,
          eid = {arXiv:1111.4246},
        pages = {arXiv:1111.4246},
          doi = {10.48550/arXiv.1111.4246},
archivePrefix = {arXiv},
       eprint = {1111.4246},
 primaryClass = {stat.CO},
       adsurl = {https://ui.adsabs.harvard.edu/abs/2011arXiv1111.4246H}
}

@ARTICLE{kirkpatrick2024,
       author = {{Kirkpatrick}, J. Davy and {Marocco}, Federico and {Gelino}, Christopher R. and {Raghu}, Yadukrishna and {Faherty}, Jacqueline K. and {Bardalez Gagliuffi}, Daniella C. and {Schurr}, Steven D. and {Apps}, Kevin and {Schneider}, Adam C. and {Meisner}, Aaron M. and {Kuchner}, Marc J. and {Caselden}, Dan and {Smart}, R.~L. and {Casewell}, S.~L. and {Raddi}, Roberto and {Kesseli}, Aurora and {Stevnbak Andersen}, Nikolaj and {Antonini}, Edoardo and {Beaulieu}, Paul and {Bickle}, Thomas P. and {Bilsing}, Martin and {Chieng}, Raymond and {Colin}, Guillaume and {Deen}, Sam and {Dereveanco}, Alexandru and {Doll}, Katharina and {Durantini Luca}, Hugo A. and {Frazer}, Anya and {Gantier}, Jean Marc and {Gramaize}, L{\'e}opold and {Grant}, Kristin and {Hamlet}, Leslie K. and {Higashimura}, Hiro and {Hyogo}, Michiharu and {Ja{\l}owiczor}, Peter A. and {Jonkeren}, Alexander and {Kabatnik}, Martin and {Kiwy}, Frank and {Martin}, David W. and {Michaels}, Marianne N. and {Pendrill}, William and {Pessanha Machado}, Celso and {Pumphrey}, Benjamin and {Rothermich}, Austin and {Russwurm}, Rebekah and {Sainio}, Arttu and {Sanchez}, John and {Sapelkin-Tambling}, Fyodor Theo and {Sch{\"u}mann}, J{\"o}rg and {Selg-Mann}, Karl and {Singh}, Harshdeep and {Stenner}, Andres and {Sun}, Guoyou and {Tanner}, Christopher and {Th{\'e}venot}, Melina and {Ventura}, Maurizio and {Voloshin}, Nikita V. and {Walla}, Jim and {W{\k{e}}dracki}, Zbigniew and {Adorno}, Jose I. and {Aganze}, Christian and {Allers}, Katelyn N. and {Brooks}, Hunter and {Burgasser}, Adam J. and {Calamari}, Emily and {Connor}, Thomas and {Costa}, Edgardo and {Eisenhardt}, Peter R. and {Gagn{\'e}}, Jonathan and {Gerasimov}, Roman and {Gonzales}, Eileen C. and {Hsu}, Chih-Chun and {Kiman}, Rocio and {Li}, Guodong and {Low}, Ryan and {Mamajek}, Eric and {Pantoja}, Blake M. and {Popinchalk}, Mark and {Rees}, Jon M. and {Stern}, Daniel and {Su{\'a}rez}, Genaro and {Theissen}, Christopher and {Tsai}, Chao-Wei and {Vos}, Johanna M. and {Zurek}, David and {The Backyard Worlds: Planet 9 Collaboration}},
        title = "{The Initial Mass Function Based on the Full-sky 20 pc Census of {\ensuremath{\sim}}3600 Stars and Brown Dwarfs}",
      journal = {\apjs},
         year = 2024,
        month = apr,
       volume = {271},
       number = {2},
          eid = {55},
        pages = {55},
          doi = {10.3847/1538-4365/ad24e2},
archivePrefix = {arXiv},
       eprint = {2312.03639},
 primaryClass = {astro-ph.SR},
       adsurl = {https://ui.adsabs.harvard.edu/abs/2024ApJS..271...55K}
}

@ARTICLE{henry2006,
       author = {{Henry}, Todd J. and {Jao}, Wei-Chun and {Subasavage}, John P. and {Beaulieu}, Thomas D. and {Ianna}, Philip A. and {Costa}, Edgardo and {M{\'e}ndez}, Ren{\'e} A.},
        title = "{The Solar Neighborhood. XVII. Parallax Results from the CTIOPI 0.9 m Program: 20 New Members of the RECONS 10 Parsec Sample}",
      journal = {\aj},
         year = 2006,
        month = dec,
       volume = {132},
       number = {6},
        pages = {2360-2371},
          doi = {10.1086/508233},
archivePrefix = {arXiv},
       eprint = {astro-ph/0608230},
 primaryClass = {astro-ph},
       adsurl = {https://ui.adsabs.harvard.edu/abs/2006AJ....132.2360H}
}

@ARTICLE{gaiadr3,
       author = {{Gaia Collaboration} and {Vallenari}, A. and {Brown}, A.~G.~A. and {Prusti}, T. and {de Bruijne}, J.~H.~J. and {Arenou}, F. and {Babusiaux}, C. and {Biermann}, M. and {Creevey}, O.~L. and {Ducourant}, C. and {Evans}, D.~W. and {Eyer}, L. and {Guerra}, R. and {Hutton}, A. and {Jordi}, C. and {Klioner}, S.~A. and {Lammers}, U.~L. and {Lindegren}, L. and {Luri}, X. and {Mignard}, F. and {Panem}, C. and {Pourbaix}, D. and {Randich}, S. and {Sartoretti}, P. and {Soubiran}, C. and {Tanga}, P. and {Walton}, N.~A. and {Bailer-Jones}, C.~A.~L. and {Bastian}, U. and {Drimmel}, R. and {Jansen}, F. and {Katz}, D. and {Lattanzi}, M.~G. and {van Leeuwen}, F. and {Bakker}, J. and {Cacciari}, C. and {Casta{\~n}eda}, J. and {De Angeli}, F. and {Fabricius}, C. and {Fouesneau}, M. and {Fr{\'e}mat}, Y. and {Galluccio}, L. and {Guerrier}, A. and {Heiter}, U. and {Masana}, E. and {Messineo}, R. and {Mowlavi}, N. and {Nicolas}, C. and {Nienartowicz}, K. and {Pailler}, F. and {Panuzzo}, P. and {Riclet}, F. and {Roux}, W. and {Seabroke}, G.~M. and {Sordo}, R. and {Th{\'e}venin}, F. and {Gracia-Abril}, G. and {Portell}, J. and {Teyssier}, D. and {Altmann}, M. and {Andrae}, R. and {Audard}, M. and {Bellas-Velidis}, I. and {Benson}, K. and {Berthier}, J. and {Blomme}, R. and {Burgess}, P.~W. and {Busonero}, D. and {Busso}, G. and {C{\'a}novas}, H. and {Carry}, B. and {Cellino}, A. and {Cheek}, N. and {Clementini}, G. and {Damerdji}, Y. and {Davidson}, M. and {de Teodoro}, P. and {Nu{\~n}ez Campos}, M. and {Delchambre}, L. and {Dell'Oro}, A. and {Esquej}, P. and {Fern{\'a}ndez-Hern{\'a}ndez}, J. and {Fraile}, E. and {Garabato}, D. and {Garc{\'\i}a-Lario}, P. and {Gosset}, E. and {Haigron}, R. and {Halbwachs}, J.-L. and {Hambly}, N.~C. and {Harrison}, D.~L. and {Hern{\'a}ndez}, J. and {Hestroffer}, D. and {Hodgkin}, S.~T. and {Holl}, B. and {Jan{\ss}en}, K. and {Jevardat de Fombelle}, G. and {Jordan}, S. and {Krone-Martins}, A. and {Lanzafame}, A.~C. and {L{\"o}ffler}, W. and {Marchal}, O. and {Marrese}, P.~M. and {Moitinho}, A. and {Muinonen}, K. and {Osborne}, P. and {Pancino}, E. and {Pauwels}, T. and {Recio-Blanco}, A. and {Reyl{\'e}}, C. and {Riello}, M. and {Rimoldini}, L. and {Roegiers}, T. and {Rybizki}, J. and {Sarro}, L.~M. and {Siopis}, C. and {Smith}, M. and {Sozzetti}, A. and {Utrilla}, E. and {van Leeuwen}, M. and {Abbas}, U. and {{\'A}brah{\'a}m}, P. and {Abreu Aramburu}, A. and {Aerts}, C. and {Aguado}, J.~J. and {Ajaj}, M. and {Aldea-Montero}, F. and {Altavilla}, G. and {{\'A}lvarez}, M.~A. and {Alves}, J. and {Anders}, F. and {Anderson}, R.~I. and {Anglada Varela}, E. and {Antoja}, T. and {Baines}, D. and {Baker}, S.~G. and {Balaguer-N{\'u}{\~n}ez}, L. and {Balbinot}, E. and {Balog}, Z. and {Barache}, C. and {Barbato}, D. and {Barros}, M. and {Barstow}, M.~A. and {Bartolom{\'e}}, S. and {Bassilana}, J.-L. and {Bauchet}, N. and {Becciani}, U. and {Bellazzini}, M. and {Berihuete}, A. and {Bernet}, M. and {Bertone}, S. and {Bianchi}, L. and {Binnenfeld}, A. and {Blanco-Cuaresma}, S. and {Blazere}, A. and {Boch}, T. and {Bombrun}, A. and {Bossini}, D. and {Bouquillon}, S. and {Bragaglia}, A. and {Bramante}, L. and {Breedt}, E. and {Bressan}, A. and {Brouillet}, N. and {Brugaletta}, E. and {Bucciarelli}, B. and {Burlacu}, A. and {Butkevich}, A.~G. and {Buzzi}, R. and {Caffau}, E. and {Cancelliere}, R. and {Cantat-Gaudin}, T. and {Carballo}, R. and {Carlucci}, T. and {Carnerero}, M.~I. and {Carrasco}, J.~M. and {Casamiquela}, L. and {Castellani}, M. and {Castro-Ginard}, A. and {Chaoul}, L. and {Charlot}, P. and {Chemin}, L. and {Chiaramida}, V. and {Chiavassa}, A. and {Chornay}, N. and {Comoretto}, G. and {Contursi}, G. and {Cooper}, W.~J. and {Cornez}, T. and {Cowell}, S. and {Crifo}, F. and {Cropper}, M. and {Crosta}, M. and {Crowley}, C. and {Dafonte}, C. and {Dapergolas}, A. and {David}, M. and {David}, P. and {de Laverny}, P. and {De Luise}, F. and {De March}, R.},
        title = "{Gaia Data Release 3. Summary of the content and survey properties}",
      journal = {\aap},
         year = 2023,
        month = jun,
       volume = {674},
          eid = {A1},
        pages = {A1},
          doi = {10.1051/0004-6361/202243940},
archivePrefix = {arXiv},
       eprint = {2208.00211},
 primaryClass = {astro-ph.GA},
       adsurl = {https://ui.adsabs.harvard.edu/abs/2023A&A...674A...1G}
}

@PROCEEDINGS{hipparcos,
        title = "{The HIPPARCOS and TYCHO catalogues. Astrometric and photometric star catalogues derived from the ESA HIPPARCOS Space Astrometry Mission}",
    booktitle = {ESA Special Publication},
         year = 1997,
       editor = {{ESA}},
       series = {ESA Special Publication},
       volume = {1200},
        month = jan,
       adsurl = {https://ui.adsabs.harvard.edu/abs/1997ESASP1200.....E}
}

@ARTICLE{DESIMW,
       author = {{Koposov}, Sergey E. and {Allende Prieto}, C. and {Cooper}, A.~P. and {Li}, T.~S. and {Beraldo e Silva}, L. and {Kim}, B. and {Carrillo}, A. and {Dey}, A. and {Manser}, C.~J. and {Nikakhtar}, F. and {Riley}, A.~H. and {Rockosi}, C. and {Valluri}, M. and {Aguilar}, J. and {Ahlen}, S. and {Bailey}, S. and {Blum}, R. and {Brooks}, D. and {Claybaugh}, T. and {Cole}, S. and {de la Macorra}, A. and {Dey}, B. and {Forero-Romero}, J.~E. and {Gazta{\~n}aga}, E. and {Guy}, J. and {Kremin}, A. and {Le Guillou}, L. and {Levi}, M.~E. and {Manera}, M. and {Meisner}, A. and {Miquel}, R. and {Moustakas}, J. and {Nie}, J. and {Palanque-Delabrouille}, N. and {Percival}, W.~J. and {Rezaie}, M. and {Rossi}, G. and {Sanchez}, E. and {Schlafly}, E.~F. and {Schubnell}, M. and {Tarl{\'e}}, G. and {Weaver}, B.~A. and {Zhou}, Z.},
        title = "{DESI Early Data Release Milky Way Survey value-added catalogue}",
      journal = {\mnras},
         year = 2024,
        month = sep,
       volume = {533},
       number = {1},
        pages = {1012-1031},
          doi = {10.1093/mnras/stae1842},
archivePrefix = {arXiv},
       eprint = {2407.06280},
 primaryClass = {astro-ph.GA},
       adsurl = {https://ui.adsabs.harvard.edu/abs/2024MNRAS.533.1012K}
}

@ARTICLE{sdssdr19,
       author = {{SDSS Collaboration} and {Adamane Pallathadka}, Gautham and {Aghakhanloo}, Mojgan and {Aird}, James and {Almeida}, Andr{\'e}s and {Amrita}, Singh and {Anders}, Friedrich and {Anderson}, Scott F. and {Arseneau}, Stefan and {Gonz{\'a}lez Avila}, Consuelo and {Aviram}, Shir and {Aydar}, Catarina and {Badenes}, Carles and {Barrera-Ballesteros}, Jorge K. and {Bauer}, Franz E. and {Behmard}, Aida and {Berg}, Michelle and {Besser}, F. and {Moni Bidin}, Christian and {Bizyaev}, Dmitry and {Blanc}, Guillermo and {Blanton}, Michael R. and {Bovy}, Jo and {Brandt}, William Nielsen and {Brownstein}, Joel R. and {Buchner}, Johannes and {Bulbul}, Esra and {Burchett}, Joseph N. and {Carigi}, Leticia and {Carlberg}, Joleen K. and {Casey}, Andrew R. and {Chakraborty}, Priyanka and {Chanam{\'e}}, Julio and {Chandra}, Vedant and {Chiappini}, Cristina and {Chilingarian}, Igor and {Comparat}, Johan and {Covey}, Kevin and {Crumpler}, Nicole and {Cunha}, Katia and {D'Onghia}, Elena and {Dai}, Xinyu and {Darling}, Jeremy and {Davis}, Megan and {De Lee}, Nathan and {Deacon}, Niall and {M{\'e}ndez Delgado}, Jos{\'e} Eduardo and {Demasi}, Sebastian and {Demianenko}, Mariia and {Demke}, Delvin and {Donor}, John and {Drory}, Niv and {Villa Durango}, Monica Alejandra and {Dwelly}, Tom and {Egorov}, Oleg and {Egorova}, Evgeniya and {El-Badry}, Kareem and {Eracleous}, Mike and {Fan}, Xiaohui and {Farr}, Emily and {Finkbeiner}, Douglas P. and {Fries}, Logan and {Frinchaboy}, Peter and {Gentile Fusillo}, Nicola Pietro and {Serrano F{\'e}lix}, Luis Daniel and {Gaensicke}, Boris and {Galligan}, Emma and {Garc{\'\i}a}, Pablo and {Gelfand}, Joseph and {Grabowski}, Katie and {Grebel}, Eva and {Green}, Paul J and {Greve}, Hannah and {Grier}, Catherine and {Griffith}, Emily and {Guetzoyan}, Paloma and {Gupta}, Pramod and {Hackshaw}, Zoe and {Hall}, Patrick B. and {Hawkins}, Keith and {Heged{\H{u}}s}, Viola and {Hekker}, Saskia and {Herbst}, T.~M. and {Hermes}, J.~J. and {Hern{\'a}ndez-Garc{\'\i}a}, Lorena and {Hiremath}, Pranavi and {Hogg}, David W and {Holtzman}, Jon and {Horne}, Keith and {Horta}, Danny and {Huang}, Yang and {Hutchinson}, Brian and {H{\"a}berle}, Maximilian and {Ibarra-Medel}, Hector Javier and {Ji}, Alexander P. and {Jofre}, Paula and {Johnson}, James W. and {Johnson}, Jennifer and {Johnston}, Evelyn J. and {Kaldor}, Mary and {Katkov}, Ivan and {Khalatyan}, Arman and {Khoperskov}, Sergey and {Klessen}, Ralf and {Kluge}, Matthias and {Koekemoer}, Anton M. and {Kollmeier}, Juna A. and {Kounkel}, Marina and {Kreckel}, Kathryn and {Krishnarao}, Dhanesh and {Krumpe}, Mirko and {Lacerna}, Ivan and {Laporte}, Chervin and {Lepine}, Sebastien and {Li}, Jing and {Liang}, Fu-Heng and {Limberg}, Guilherme and {Liu}, Xin and {Loebman}, Sarah and {Long}, Knox and {Lu}, Yuxi and {Lucey}, Madeline and {Lugo-Aranda}, Alejandra Z. and {Mart{\'\i}nez Martinez-Aldama}, Mary Loli and {McKinnon}, Kevin and {Medan}, Ilija and {Merloni}, Andrea and {Morrison}, Sean and {Myers}, Natalie and {M{\'e}sz{\'a}ros}, Szabolcs and {M{\"u}ller-Horn}, Johanna and {Nepal}, Samir and {Ness}, Melissa and {Nidever}, David and {Nitschelm}, Christian and {Oravetz}, Audrey and {Otto}, Jonah and {Pan}, Kaike and {P{\'e}rez Paolino}, Facundo and {Negrete Pe{\~n}aloza}, Castalia Alenka and {Pinsonneault}, Marc and {Taghizadeh Popp}, Manuchehr and {Price-Whelan}, Adrian and {Pulatova}, Nadiia and {Queiroz}, Anna Barbara and {Raddick}, Jordan and {Rankine}, Amy and {Rix}, Hans-Walter and {Rom{\'a}n-Z{\'u}{\~n}iga}, Carlos and {Fern{\'a}ndez Rosso}, Daniela and {Runnoe}, Jessie and {Mahmud Saad}, Serat and {Salvato}, Mara and {Sanchez}, Sebastian F. and {Sattler}, Natascha and {Saydjari}, Andrew and {Sayres}, Conor and {Schlaufman}, Kevin and {Schneider}, Donald P. and {Schwope}, Axel and {Seaton}, Lucas M. and {Seeburger}, Rhys and {Serna}, Javier and {Sharma}, Sanjib and {Shen}, Yue and {Sinha}, Amaya and {Sizemore}, Brian and {Sniegowska}, Marzena and {Song}, Yingyi and {Souto}, Diogo and {Stassun}, Keivan and {Steinmetz}, Matthias and {Stone}, Zachary and {Stone-Martinez}, Alexander and {Stringfellow}, Guy S. and {Mata S{\'a}nchez}, Aurora and {S{\'a}nchez-Gallego}, Jos{\'e} and {Tan}, Jonathan and {Tayar}, Jamie and {Thai}, Riley and {Thakar}, Ani and {Thibodeaux}, Pierre and {Ting}, Yuan-Sen and {Tkachenko}, Andrew and {Trakhtenbrot}, Benny and {Fernandez Trincado}, Jose G. and {Troup}, Nicholas and {Trump}, Jonathan R. and {Ulloa}, Natalie and {Van der Marel}, Roeland P. and {Vera}, Pablo and {Villanova}, Sandro and {Villase{\~n}or}, Jaime and {Wang}, Ji and {Way}, Zachary and {Weijmans}, Anne-Marie and {Wheeler}, Adam and {Wilson}, John C. and {Wofford}, Aida and {Wong}, Tony},
        title = "{The Nineteenth Data Release of the Sloan Digital Sky Survey}",
      journal = {arXiv e-prints},
         year = 2025,
        month = jul,
          eid = {arXiv:2507.07093},
        pages = {arXiv:2507.07093},
          doi = {10.48550/arXiv.2507.07093},
archivePrefix = {arXiv},
       eprint = {2507.07093},
 primaryClass = {astro-ph.GA},
       adsurl = {https://ui.adsabs.harvard.edu/abs/2025arXiv250707093S}
}

@ARTICLE{slam,
       author = {{Qiu}, Dan and {Johnson}, Jennifer A. and {Liu}, Chao and {Souto}, Diogo and {Medan}, Ilija and {Stringfellow}, Guy S. and {Way}, Zachary and {Ting}, Yuan-sen and {Casey}, Andrew R. and {Rojas-Ayala}, B{\'a}rbara and {L{\'o}pez-Valdivia}, Ricardo and {Song}, Ying-Yi and {Zhang}, Bo and {Li}, Jiadong and {Behmard}, Aida and {M{\'e}sz{\'a}ros}, Szabolcs and {Stassun}, Keivan G. and {Fern{\'a}ndez-Trincado}, Jos{\'e} G.},
        title = "{Stellar Parameters of BOSS M dwarfs in SDSS-V DR19}",
      journal = {arXiv e-prints},
         year = 2025,
        month = nov,
          eid = {arXiv:2511.20005},
        pages = {arXiv:2511.20005},
          doi = {10.48550/arXiv.2511.20005},
archivePrefix = {arXiv},
       eprint = {2511.20005},
 primaryClass = {astro-ph.SR},
       adsurl = {https://ui.adsabs.harvard.edu/abs/2025arXiv251120005Q}
}

@article{way2026,
doi = {10.3847/1538-3881/ae48ae},
url = {https://doi.org/10.3847/1538-3881/ae48ae},
year = {2026},
month = {mar},
publisher = {The American Astronomical Society},
volume = {171},
number = {4},
pages = {252},
author = {Way, Zachary and Lépine, Sébastien and Gagné, Jonathan and Medan, Ilija},
title = {21,864 Unresolved, Low-mass Binaries Identified via Their Overluminosity in Gaia Data Release 3 and a Catalog of 347,440 Systems within 100 pc of the Sun},
journal = {The Astronomical Journal}
}

@article{Bressan2012,
  author  = {Bressan, A. and Marigo, P. and Girardi, L. and et al.},
  title   = {PARSEC: stellar tracks and isochrones with the PAdova and TRieste Stellar Evolution Code},
  journal = {Monthly Notices of the Royal Astronomical Society},
  volume  = {427},
  pages   = {127--145},
  year    = {2012}
}

@article{Choi2016,
  author  = {Choi, J. and Dotter, A. and Conroy, C. and et al.},
  title   = {Mesa Isochrones and Stellar Tracks (MIST). I. Solar-scaled Models},
  journal = {The Astrophysical Journal},
  volume  = {823},
  pages   = {102},
  year    = {2016}
}

@ARTICLE{elbadry2021,
       author = {{El-Badry}, Kareem and {Rix}, Hans-Walter and {Heintz}, Tyler M.},
        title = "{A million binaries from Gaia eDR3: sample selection and validation of Gaia parallax uncertainties}",
      journal = {\mnras},
         year = 2021,
        month = sep,
       volume = {506},
       number = {2},
        pages = {2269-2295},
          doi = {10.1093/mnras/stab323},
archivePrefix = {arXiv},
       eprint = {2101.05282},
 primaryClass = {astro-ph.SR},
       adsurl = {https://ui.adsabs.harvard.edu/abs/2021MNRAS.506.2269E}
}

@ARTICLE{mann2019,
       author = {{Mann}, Andrew W. and {Dupuy}, Trent and {Kraus}, Adam L. and {Gaidos}, Eric and {Ansdell}, Megan and {Ireland}, Michael and {Rizzuto}, Aaron C. and {Hung}, Chao-Ling and {Dittmann}, Jason and {Factor}, Samuel and {Feiden}, Gregory and {Martinez}, Raquel A. and {Ru{\'\i}z-Rodr{\'\i}guez}, Dary and {Thao}, Pa Chia},
        title = "{How to Constrain Your M Dwarf. II. The Mass-Luminosity-Metallicity Relation from 0.075 to 0.70 Solar Masses}",
      journal = {\apj},
         year = 2019,
        month = jan,
       volume = {871},
       number = {1},
          eid = {63},
        pages = {63},
          doi = {10.3847/1538-4357/aaf3bc},
archivePrefix = {arXiv},
       eprint = {1811.06938},
 primaryClass = {astro-ph.SR},
       adsurl = {https://ui.adsabs.harvard.edu/abs/2019ApJ...871...63M}
}

@ARTICLE{li2023,
       author = {{Li}, Jiadong and {Liu}, Chao and {Zhang}, Zhi-Yu and {Tian}, Hao and {Fu}, Xiaoting and {Li}, Jiao and {Yan}, Zhi-Qiang},
        title = "{Stellar initial mass function varies with metallicity and time}",
      journal = {\nat},
         year = 2023,
        month = jan,
       volume = {613},
       number = {7944},
        pages = {460-462},
          doi = {10.1038/s41586-022-05488-1},
archivePrefix = {arXiv},
       eprint = {2301.07029},
 primaryClass = {astro-ph.GA},
       adsurl = {https://ui.adsabs.harvard.edu/abs/2023Natur.613..460L}
}

@ARTICLE{CNS5,
       author = {{Golovin}, Alex and {Reffert}, Sabine and {Just}, Andreas and {Jordan}, Stefan and {Vani}, Akash and {Jahrei{\ss}}, Hartmut},
        title = "{The Fifth Catalogue of Nearby Stars (CNS5)}",
      journal = {\aap},
         year = 2023,
        month = feb,
       volume = {670},
          eid = {A19},
        pages = {A19},
          doi = {10.1051/0004-6361/202244250},
archivePrefix = {arXiv},
       eprint = {2211.01449},
 primaryClass = {astro-ph.SR},
       adsurl = {https://ui.adsabs.harvard.edu/abs/2023A&A...670A..19G}
}

@ARTICLE{kroupa2001,
       author = {{Kroupa}, Pavel},
        title = "{On the variation of the initial mass function}",
      journal = {\mnras},
         year = 2001,
        month = apr,
       volume = {322},
       number = {2},
        pages = {231-246},
          doi = {10.1046/j.1365-8711.2001.04022.x},
archivePrefix = {arXiv},
       eprint = {astro-ph/0009005},
 primaryClass = {astro-ph},
       adsurl = {https://ui.adsabs.harvard.edu/abs/2001MNRAS.322..231K}
}

@article{kullback1951information,
  title={On information and sufficiency},
  author={Kullback, Solomon and Leibler, Richard A},
  journal={The annals of mathematical statistics},
  volume={22},
  number={1},
  pages={79--86},
  year={1951},
  publisher={JSTOR}
}

@ARTICLE{Riello2021,
       author = {{Riello}, M. and {De Angeli}, F. and {Evans}, D.~W. and {Montegriffo}, P. and {Carrasco}, J.~M. and {Busso}, G. and {Palaversa}, L. and {Burgess}, P.~W. and {Diener}, C. and {Davidson}, M. and {Rowell}, N. and {Fabricius}, C. and {Jordi}, C. and {Bellazzini}, M. and {Pancino}, E. and {Harrison}, D.~L. and {Cacciari}, C. and {van Leeuwen}, F. and {Hambly}, N.~C. and {Hodgkin}, S.~T. and {Osborne}, P.~J. and {Altavilla}, G. and {Barstow}, M.~A. and {Brown}, A.~G.~A. and {Castellani}, M. and {Cowell}, S. and {De Luise}, F. and {Gilmore}, G. and {Giuffrida}, G. and {Hidalgo}, S. and {Holland}, G. and {Marinoni}, S. and {Pagani}, C. and {Piersimoni}, A.~M. and {Pulone}, L. and {Ragaini}, S. and {Rainer}, M. and {Richards}, P.~J. and {Sanna}, N. and {Walton}, N.~A. and {Weiler}, M. and {Yoldas}, A.},
        title = "{Gaia Early Data Release 3. Photometric content and validation}",
      journal = {\aap},
         year = 2021,
        month = may,
       volume = {649},
          eid = {A3},
        pages = {A3},
          doi = {10.1051/0004-6361/202039587},
archivePrefix = {arXiv},
       eprint = {2012.01916},
 primaryClass = {astro-ph.IM},
       adsurl = {https://ui.adsabs.harvard.edu/abs/2021A&A...649A...3R}
}

@ARTICLE{Fabricius2021,
       author = {{Fabricius}, C. and {Luri}, X. and {Arenou}, F. and {Babusiaux}, C. and {Helmi}, A. and {Muraveva}, T. and {Reyl{\'e}}, C. and {Spoto}, F. and {Vallenari}, A. and {Antoja}, T. and {Balbinot}, E. and {Barache}, C. and {Bauchet}, N. and {Bragaglia}, A. and {Busonero}, D. and {Cantat-Gaudin}, T. and {Carrasco}, J.~M. and {Diakit{\'e}}, S. and {Fabrizio}, M. and {Figueras}, F. and {Garcia-Gutierrez}, A. and {Garofalo}, A. and {Jordi}, C. and {Kervella}, P. and {Khanna}, S. and {Leclerc}, N. and {Licata}, E. and {Lambert}, S. and {Marrese}, P.~M. and {Masip}, A. and {Ramos}, P. and {Robichon}, N. and {Robin}, A.~C. and {Romero-G{\'o}mez}, M. and {Rubele}, S. and {Weiler}, M.},
        title = "{Gaia Early Data Release 3. Catalogue validation}",
      journal = {\aap},
         year = 2021,
        month = may,
       volume = {649},
          eid = {A5},
        pages = {A5},
          doi = {10.1051/0004-6361/202039834},
archivePrefix = {arXiv},
       eprint = {2012.06242},
 primaryClass = {astro-ph.GA},
       adsurl = {https://ui.adsabs.harvard.edu/abs/2021A&A...649A...5F}
}

@ARTICLE{Karim2017,
       author = {{Karim}, T. and {Mamajek}, Eric E.},
        title = "{Revised geometric estimates of the North Galactic Pole and the Sun's height above the Galactic mid-plane}",
      journal = {\mnras},
         year = 2017,
        month = feb,
       volume = {465},
       number = {1},
        pages = {472-481},
          doi = {10.1093/mnras/stw2772},
archivePrefix = {arXiv},
       eprint = {1610.08125},
 primaryClass = {astro-ph.SR},
       adsurl = {https://ui.adsabs.harvard.edu/abs/2017MNRAS.465..472K}
}

@software{snc_sf_zenodo,
  author       = {Ilija Medan},
  title        = {imedan/snc\_sf: 1.1.1},
  month        = jul,
  year         = 2026,
  publisher    = {Zenodo},
  version      = {1.1.1},
  doi          = {10.5281/zenodo.21285596},
  url          = {https://doi.org/10.5281/zenodo.21285596},
}
\bibliographystyle{aasjournal}

\appendix

\section{Selection Function Code Example}\label{app:example}

Below, we will go through a simple example of how to use the code associated with this work\footnote{\url{https://github.com/imedan/snc_sf/}; \url{https://doi.org/10.5281/zenodo.21285596}}\added{ \citep{snc_sf_zenodo}}. For this example, we will only consider the APOGEE SNC stars from DR19. The SDSS-V data needed from this tutorial can be downloaded directly from the SAS\footnote{\url{https://data.sdss.org/sas/dr19/spectro/astra/0.6.0/summary/astraAllStarASPCAP-0.6.0.fits.gz}}. Below we show how to open the resulting file with the APOGEE data, filter on SNC targets and save the resulting data for later.
\begin{lstlisting}[language=Python]
from astropy.table import Table, unique
from sdss_semaphore.targeting import TargetingFlags

# open the DR19 file with the APOGEE data
dr19 = Table.read('astraAllStarASPCAP-0.6.0.fits.gz', hdu=2)

# use semaphore to identify stars in the SNC carton
flags = TargetingFlags(dr19['sdss5_target_flags'])
snc = flags.in_carton_name('mwm_snc_100pc')

# save the data
snc_100pc = unique(dr19[snc & (dr19['plx'] > 0)],
                   keys='gaia_dr3_source_id',
                   keep='first')
snc_100pc.rename_column('gaia_dr3_source_id', 'source_id')
snc_100pc[['source_id', 'ra', 'dec', 'fe_h', 'e_fe_h']].write('obs_100pc_edr3.csv',
                                                              format='csv',
                                                              overwrite=True)
\end{lstlisting}
This saved file will serve as the basis of the observed data for our code. It is important to note that the GCNS was run with eDR3 data, so it is important that the required columns for the selection function match eDR3. The code automatically matches these for you based on the Gaia \texttt{source\_id}. It will replace columns already present in the above file, like \texttt{ra}, if they exist though.

With the above file save, we can now run our selection function code! First, we need to calculate the selection function for this subset of 100 pc objects, as described in Section \ref{sec:select_func}. As a reminder, this is based on all of the SNC data considered, which is the DR19 100 pc sample with an APOGEE spectrum. Below we show how the object is initialized. When initialized, it will automatically calculate the selection function. Additionally, if this is the first time running the code, it will automatically download the necessary GCNS data files.
\begin{lstlisting}[language=Python]
# load all needed packages and functions
from snc_sf.selection_function import SNCSelectionFunction

import numpy as np
import polars as pl
import matplotlib.pylab as plt
from matplotlib.colors import LogNorm
import jax
jax.config.update("jax_enable_x64", False)

# below we configure the binning for the HR diagram for the subpopulation
# and the selection function, respectively
MG_bin_list = [0, 20, 0.25]  # [min, max, delta]

sf_bins={'healpix': 3,                   # order
         'phot_g_mean_mag': [0, 22, 1],  # [min, max, delta]
         'bp_rp': [-0.4, 5.25, 0.15]}    # [min, max, delta]

# name of the file with the query result from above
data_file = 'appendix_example.csv'

# initilize the object
mean = False
sf = SNCSelectionFunction(data_file, sf_bins, MG_bin_list, mean=mean)
\end{lstlisting}
Now the above objects includes the selection function! Within our object, we have two Polars\footnote{\url{https://pola.rs}} dataframes; one for the GCNS and one for the data. These are accessed via \texttt{sf.gcns} and \texttt{sf.data}, respectively. To access the dataframe with the $k$ and $n$ values for each bin in the selection function, this can be found in the attribute \texttt{sf.subsamp}. These $k$ and $n$ values are also joined to the GCNS and DR19 data within their respective dataframes.

Next, we will select our subpopulation to then get the subpopulation probabilities. For this example, our subpopulation will stars observed with the SNC with $[Fe/H] > -0.5$, The below code shows how to filter the subpopulation and then run the MCMC to sample the posteriors of the subpopulation probabilities, $p_{\mathsf{sub}, k}$.
\begin{lstlisting}[language=Python]
# initiallly calculate effective selection factor to get binning for GCNS
sf.evalutate_Ajk(weight_volume=False)

# filter dataset for metal-rich stars
filter_data = sf.data.filter(pl.col('fe_h') > -0.5)

# run the MCMC and get the subpopulation probability samples
p_samples, ev_valid = sf.forward_model(filter_data,
                                       num_warmup=1000,
                                       num_samples=2000,
                                       num_chains=1)
\end{lstlisting}

\begin{figure*}
	\centering
	\includegraphics[width=0.3\textwidth]{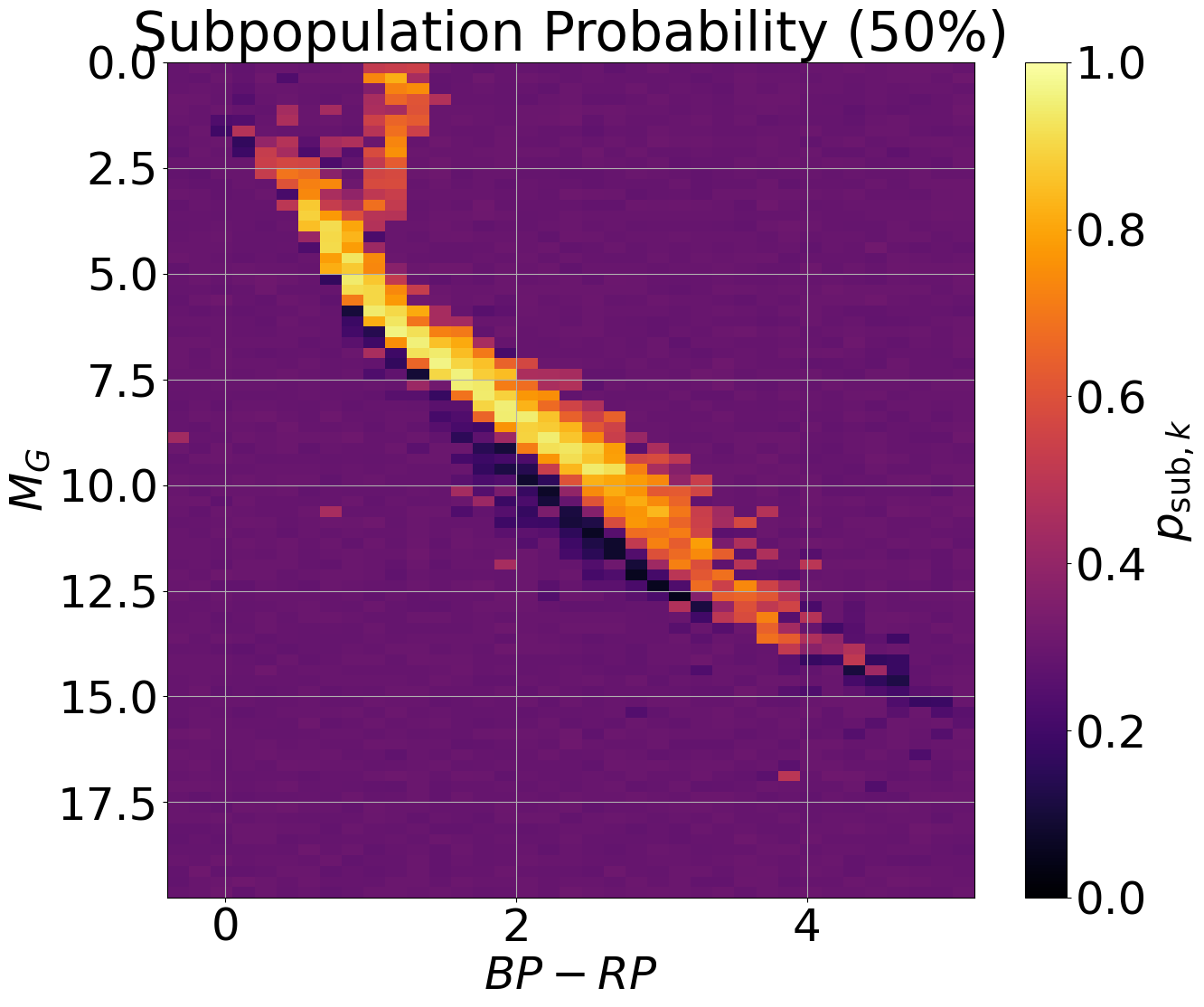}
    \includegraphics[width=0.3\textwidth]{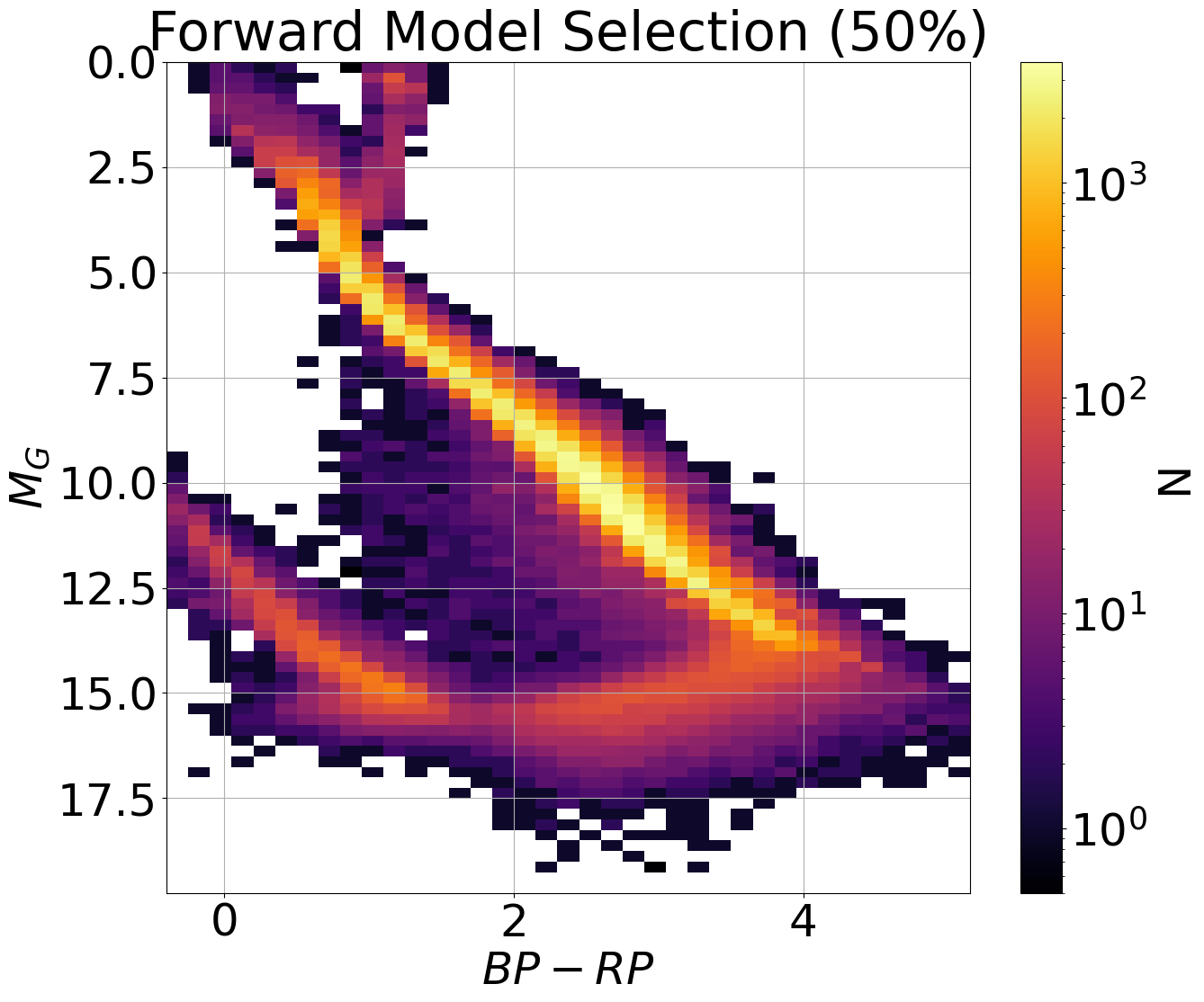}
    \includegraphics[width=0.3\textwidth]{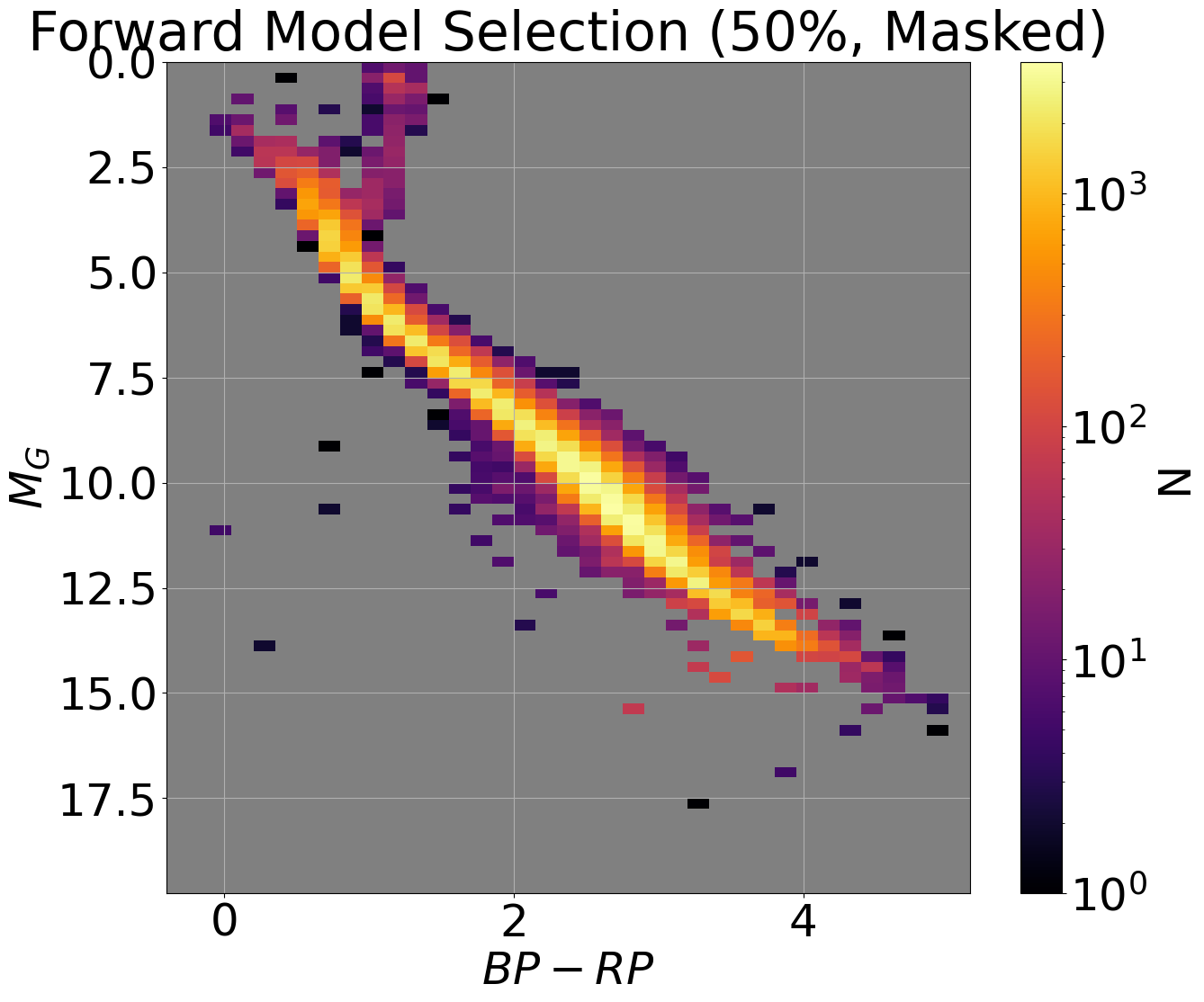}
	\caption{HR diagram of the 50th percentile of the samples of the subpopulation probabilities (left panel) and 50th percentile of the GCNS counts (middle panel) for stars with $[Fe/H] > -0.5$. The right panel shows the same HR diagram for the GCNS counts, but with regions masked with grey where the subpopulation probability posterior does not deviate significantly from the prior. The forward model correctly identifies metal-rich stars as being redder on the HR diagram.}
	\label{fig:appendix_ex}
\end{figure*}

Finally, we show a few examples on how to use these results to plot the subpopulation in the GCNS. First, we will show how to plot the 50th percentile of the samples of the subpopulation probabilities across the HR diagram (Figure \ref{fig:appendix_ex}, left panel):
\begin{lstlisting}[language=Python]
# have the bins for plotting
bp_rp_bins = np.arange(*sf.sf_bins['bp_rp'])
MG_bins = np.arange(*MG_bin_list)

# numbers in bin for reshaping the flattened p_samples
n_bp_rp = len(bp_rp_bins) - 1
n_mg = len(MG_bins) - 1

# get the 50th percentile of the samples in 2D array
p_50perc = np.nanpercentile(p_samples.reshape((-1, n_bp_rp, n_mg)), 50, axis=0)

# plot the results
f, ax1 = plt.subplots(1, 1, figsize=(12 * 1.2, 10 * 1.2))
dens = ax1.imshow(p_50perc.T,
                  origin='lower', aspect='auto',
                  extent=(bp_rp_bins.min(), bp_rp_bins.max(), MG_bins.min(), MG_bins.max()),
                  vmin=0, vmax=1, cmap='inferno')
ax1.set_xlabel(r'$BP-RP$')
ax1.set_ylabel(r'$M_G$')
ax1.invert_yaxis()
ax1.grid()
ax1.set_title('Subpopulation Probability (50%)')
plt.colorbar(dens, ax=ax1, label=r'$p_{\mathsf{sub}, k}$')
plt.show()
\end{lstlisting}
We can also plot the 50th percentile of the counts across the HR diagram for this subpopulation (Figure \ref{fig:appendix_ex}, right panel):
\begin{lstlisting}[language=Python]
# random seed to reproduce results
RNG = np.random.default_rng(666)

# get the total number of stars in the GCNS across the HR diagram
gcns_filter = sf.gcns.filter(pl.Series(sf.gcns_valid))
Ntot, _, _ = np.histogram2d(gcns_filter['bp_rp'].to_numpy(), gcns_filter['MG'].to_numpy(),
                            bins=[bp_rp_bins, MG_bins])

# based on p_samples, draw number in subpopulation
Nsub = RNG.binomial(Ntot.astype(int), p_samples.reshape((-1, n_bp_rp, n_mg)))

# get 50th percentile
Nsub_50perc = np.nanpercentile(Nsub, 50, axis=0)

# plot the results
f, ax1 = plt.subplots(1, 1, figsize=(12 * 1.2, 10 * 1.2))
dens = ax1.imshow(Nsub_50perc.T,
                  origin='lower', aspect='auto',
                  extent=(bp_rp_bins.min(), bp_rp_bins.max(), MG_bins.min(), MG_bins.max()),
                  norm=LogNorm(), cmap='inferno')
ax1.set_xlabel(r'$BP-RP$')
ax1.set_ylabel(r'$M_G$')
ax1.invert_yaxis()
ax1.grid()
ax1.set_title('Forward Model Selection (50%)')
plt.colorbar(dens, ax=ax1, label='N')
plt.show()
\end{lstlisting}

Finally, the code provides ways to flag subpopulation probability 
posteriors that are the same as the prior assumption. Below we show how 
to output these flags and identify the "bad" regions across the HR 
diagram.
\begin{lstlisting}[language=Python]
dist_change, kl_vals, kl_mean, kl_std, sample_mean, sample_var, prior_mean, prior_var = sf.check_posterior_samples(
    filter_data, p_samples)


# mask out regions returning the prior
Nsub_50perc[~dist_change.reshape((n_bp_rp, n_mg))] = np.nan

# plot the results
f, ax1 = plt.subplots(1, 1, figsize=(12 * 1.2, 10 * 1.2))

# colormap to grey out nan regions
cmap = plt.cm.inferno.copy()
cmap.set_bad('grey')

dens = ax1.imshow(Nsub_50perc.T,
                  origin='lower', aspect='auto',
                  extent=(bp_rp_bins.min(), bp_rp_bins.max(), MG_bins.min(), MG_bins.max()),
                  norm=LogNorm(), cmap=cmap)
ax1.set_xlabel(r'$BP-RP$')
ax1.set_ylabel(r'$M_G$')
ax1.invert_yaxis()
ax1.grid()
ax1.set_title('Forward Model Selection (50%, Masked)')
plt.colorbar(dens, ax=ax1, label='N')
plt.show()
\end{lstlisting}
These plots demonstrate that the model correctly identifies more 
metal-rich stars as being redder on the HR diagram. As a note, these 
plots only show the 50th percentile. With the variable \texttt{p\_samples},
 you have access to the full posterior of the subpopulation probability.
 Similar to the results in the main paper, with this you will find that 
under-sampled regions are less constrained and will typically return the
 prior for the subpopulation probability, as indicated by the masked, 
grey regions. For example, this occurs for the entire white dwarf 
sequence in this instance.

\section{ASPCAP Metallicity Correction}\label{app:metal_corr}

\added{To use the ASPCAP and SLAM metallicities together for the analysis in Section \ref{sec:mf}, corrections had to be made to the ASPCAP metallicities at lower temperatures. To determine the level of correction, we utilized the wide binaries from \cite{elbadry2021}. We select primaries with $M_G < 7$ and secondaries with $M_G > 7$. We also only choose pairs where the ASPCAP parameters have \texttt{flag\_warn = False}, \texttt{flag\_bad = False} and \texttt{e\_fe\_h = < 0.1}. For this analysis, we used the ASPCAP parameters from the internal IPL-4, which was run using tag 0.8.0 of astra and covers up to MJD $= 60708$.}

\begin{figure*}
	\centering
	\includegraphics[width=0.34\textwidth]{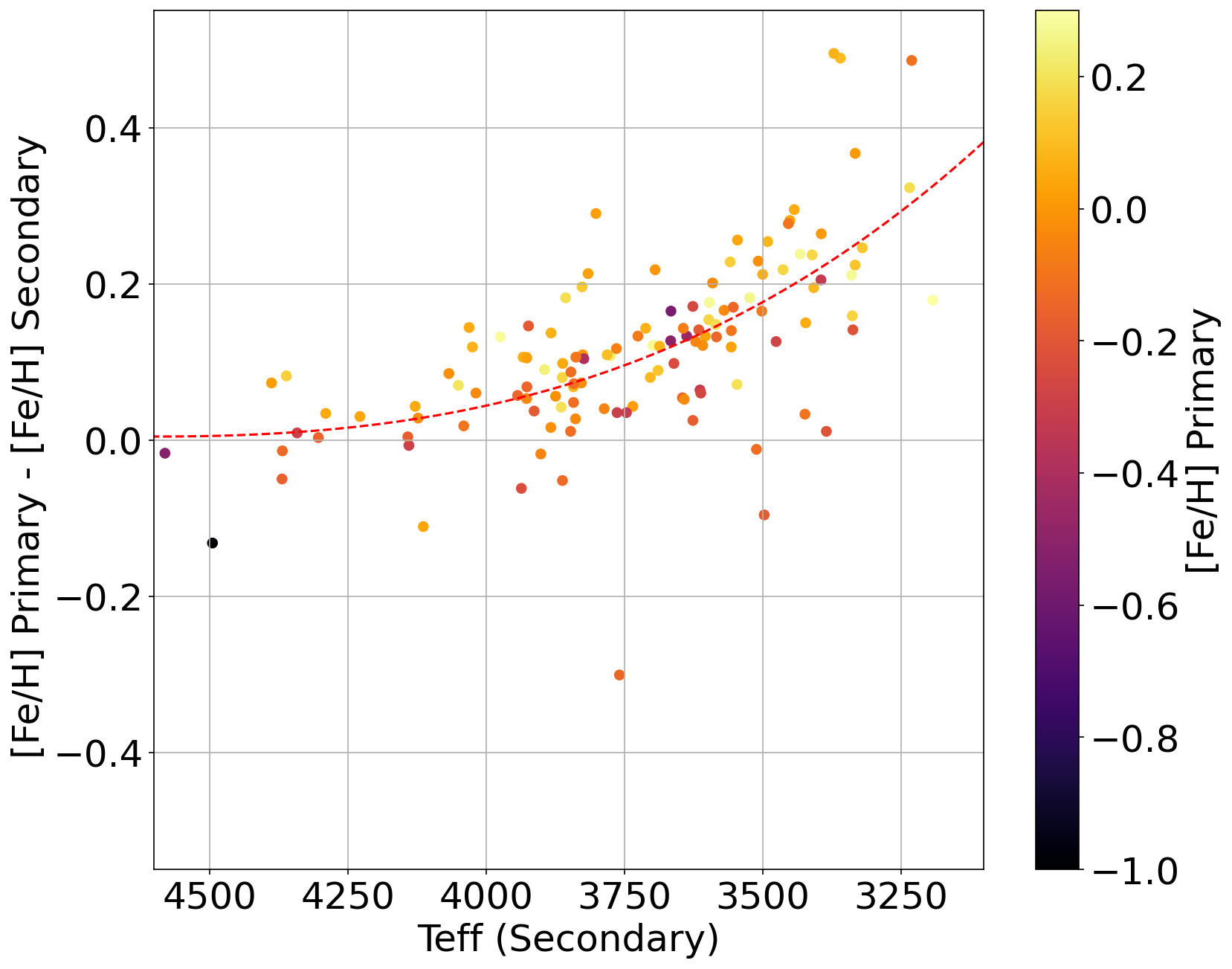}
    \includegraphics[width=0.6\textwidth]{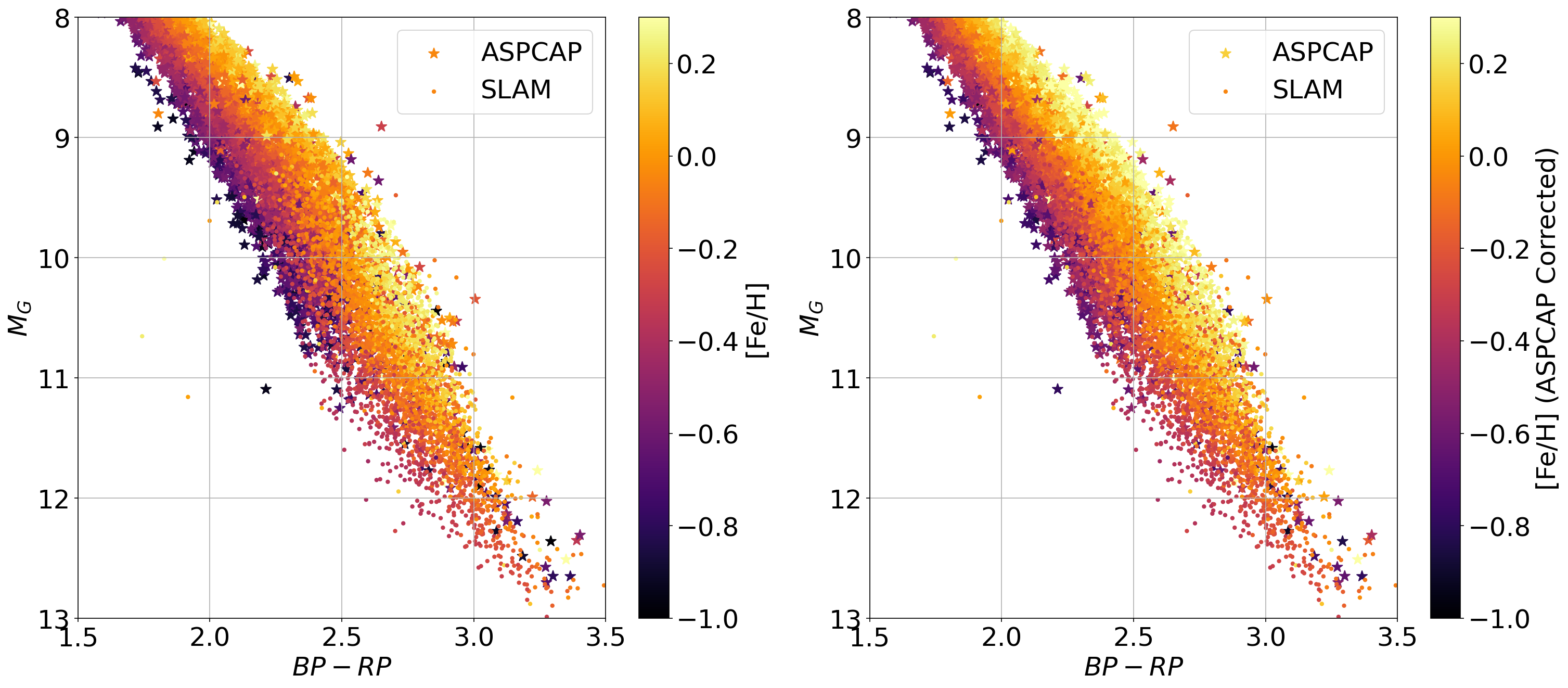}
	\caption{Left panel shows the difference in ASPCAP metallicity for wide binaries as a function of temperature of the secondary. A clear offset is observed that is well fit by a second degree polynomial (eq. \ref{eq:aspcap_metal}). The middle and right panels shows the the HR diagram of the ASPCAP and SLAM sample before and after this correction is applied, respectively.}
	\label{fig:slam_aspcap}
\end{figure*}

\added{This results in the 129 systems shown in the left panel of Figure \ref{fig:slam_aspcap}. This shows the difference in metallicity as a function of the temperature of the secondary. Fitting a second degree polynomial to these differences results in:
\begin{equation}\label{eq:aspcap_metal}
\begin{split}
\delta [\mathrm{Fe/H}] ={}& -6.095 \times 10^{-11} \, T_{\mathrm{eff}}^3
+ 9.164 \times 10^{-7} \, T_{\mathrm{eff}}^2 \\
& - 4.564 \times 10^{-3} \, T_{\mathrm{eff}} + 7.538
\end{split}
\end{equation}
We use the above to correct the ASPCAP metallicities for $T_{\rm eff} < 4600$ K in the analysis in Section \ref{sec:mf}. The middle and right panels of Figure \ref{fig:slam_aspcap} shows the HR diagram before and after this correction is applied, respectively. It is clear that before the correction, the metallicities of the ASPCAP stars in the overlap region with SLAM are too metal-poor. After the correction, the two datasets are compatible in this overlap region.}

\end{document}